# A new technique for tokamak edge density measurement based on microwave interferometer

Mariia Usoltceva[1], Stéphane Heuraux[2], Ildar Khabibullin[3] and Helmut Faugel[1]

[1] Max-Planck-Institut für Plasmaphysik, Boltzmannstr. 2, 85748 Garching, Germany
[2] Université de Lorraine, CNRS Institut Jean Lamour BP50840, F-54011 Nancy, France
[3] Max-Planck-Institut für Astrophysik, Karl-Schwarzschild-Straße 1, 85748 Garching, Germany

E-mail: mariia.usoltceva@ipp.mpg.de

**Abstract**

Novel approach for density measurements at the edge of a hot plasma device is presented – Microwave Interferometer in the Limiter Shadow (MILS). The diagnostic technique is based on measuring the change in phase and power of a microwave beam passing tangentially through the edge plasma. The wave propagation involves varying combinations of refraction, phase change and further interference of the beam fractions. A 3D model is constructed as a synthetic diagnostic for MILS and allows exploring this broad range of wave propagation regimes. The diagnostic parameters, such as its dimensions, frequency and configuration of the emitter and receiver antennas, should be balanced to meet the target range and location of measurements. It can be therefore adjusted for various conditions and here the diagnostic concept is evaluated on a chosen example, which was taken as suitable to cover densities of $\sim 10^{15}$–$10^{19}$ m$^{-3}$ on the edge of the ASDEX Upgrade tokamak. Based on a density profile with fixed radial shape, appropriate for experimental density approximation, a database of syntethic diagnostic measurements is built. The developed genetic algorithm genMILS of density profile reconstruction using the constructed database results in quite low numerical error. It is estimated as $\sim$ 5–15 % for density $\geq 10^{17}$ m$^{-3}$. Therefore, the new diagnostic technique (with dedicated data processing algorithm) has a large potential in practical applications in a wide range of densities, with low numerical error, so the total error and the density estimation accuracy is expected to be defined by experimental uncertainties.

## 1. Introduction

Plasma density of the order $\sim \leq 10^{18}$ m$^{-3}$ is typically poorly diagnosed in the conditions of a tokamak edge. Firstly, many diagnostics, which measure in that region, aim their studies at higher densities ($10^{18}$–$10^{19}$ m$^{-3}$) and are not designed to have low enough error bars for small densities. Secondly, the unstable nature of the scrape-off-layer (SOL) plasma makes it difficult to capture correct values, when their temporal fluctuations are of comparable order of magnitude as the values themselves, so this needs to be taken into account for the interpretation of measurements. In these circumstances, a diagnostic targeted specifically at the lower end of tokamak edge densities can bring valuable addition to the description of the tokamak SOL plasma.

Microwave interferometry is a very common tool for measuring plasma density, due to its robustness, wide coverage of the density range, non-intrusive measurement approach and low susceptibility to external impacts. In fusion-related plasma physics, many small- and medium-size tokamaks (JET [1], DIII-D [2] COMPASS [3], COMPASS-U [4], KSTAR [5], SUNIST [6]), stellarators (TJ-II [7], Uragan-3M [8], Uragan-2M [9]) and other devices (Aline [10], LDX [11], a capacitive RF plasma device [12]) use (or used) interferometric diagnositcs in the microwave range for measuring density in the central region of the plasma.

Measurements of the SOL density of a tokamak/stellarator with microwave techniques are usually based on the reflectometry, while the interferometric approach has not, to our knowledge, been previously attempted. Here we suggest a new measurement technique for edge densities, based on a microwave interferometer. The concept is investigated with numerical tools in this paper and the resulting data interpretation method is intended to be applied in practice on ASDEX Upgrade tokamak (experimental application is beyond the scope and possible reasonable size of this paper).

The diagnostic technique, refered by the name Microwave Interferometer in the Limiter Shadow (MILS), consists of sending a microwave signal tangentially through the edge plasma and by this measuring electron density in the vicinity of a limiter. The refraction, which in usual interferometer systems is undesirable and is minimized by directing the beam through the plasma center, is used as an advantage in MILS, by measuring the amplitude of the partially refracted signal with respect to the signal received in vacuum. Therefore, the diagnostic relies on two measured quantities, phase shift and power alteration in plasma. In order to exctract information about the plasma density from measuring these two quantities, a method for data interpretation has to be developed from scratch, the common methods being inapplicable to this kind of interferometer.

The paper is structured as follows. In the next section, we explain the choice of MILS parameters, used in this paper to study the capabilities of this measurement technique. In Section 3, the construction of the synthetic diagnostic model is explained, which is used for building a database for data interpretation, and in Section 4 the definition of the density profile in the region of interest is discussed. The model is tested on simple cases, which could be compared to analytical calculations, as shown in Section 5. The database for the density profile reconstruction is built with profiles from Section 4, which are relevant for experimental tokamak conditions, and it is described in Section 6. An analysis is done to determine the localization of the obtained density profiles, as presented in Section 7, and it is followed by examples of density reconstruction in Section 8. A discussion of the developed method of MILS data interpretation and of the new measurement technique in general is given in Section 9, followed by conclusions. The appendix contains details on the step-by-step model optimization and on the settings of the final model used in the study.

## 2. Choice of diagnostic parameters

The general concept of the diagnostic method is presented here and its capabilities are investigated on an example, which has practical interest as an application to ASDEX Upgrade tokamak. The MILS parameters are chosen as suitable to cover a range of density and of radial location, relevant to ASDEX Upgrade experimental conditions. The diagnostic concept is not limited to the parameters of the employed example.

The parameters which can be varied are the interferometer axis length $d$ and the frequency $f$. Furthermore, the horn antennas can be directed towards each other, on one line, as in conventional interferometers, or they can have another, more complex, configuration. The simplest configuration with both antennas on the same axis is considered in this study (Fig. 1).

From the general interferometer theory, which gives a zero-order approximation for MILS, in order to measure denisties $n_{min} \sim 10^{15}$ m$^{-3}$ with the possible sensitivity of the phase shift $\Delta\varphi_{min} \sim 1°$, the needed frequency is $f = \frac{n_{min} d e^2}{4\pi c \varepsilon_0 m_e \Delta\varphi_{min}}$ (see formulas in Section 3.3 for detailed derivation). The horns should be separated by at least the value of the far-field distance $d_{FF} = \frac{2D^2 f}{c}$, where $D$ is the microwave antenna aperture size, in order to have the wavefront shape close to the plane wave at the receiver location.

With $d \sim 1$ m, a realistic order of length of the interferometer axis inside a medium-size tokamak, the frequency falls in the range of $\sim 50$ GHz, which is a reasonable value for practical applications. In ASDEX Upgrade, this range of frequencies is utilized for reflectometry [13]. The size of microwave horn antennas used by this reflectometer is $D = 27$ mm, typical for this density range. An antenna with this size can provide narrow enough beam width of $\sim 20°$, in order to not scatter the radiated beam over a large area, but keep it focused close to the axis. For such antenna and at $f \sim 50$ GHz, $d_{FF}$ is $\sim 0.25$ m.

A smaller axis length would result in the signal being collected from shorter radial extent of the plasma, which narrows the boundaries of the measurement localization and can make the density reconstruction in the shorter radial region more precise. At the same time, shorter $d$ leads to larger $n_{min}$. In summary, the MILS parameters can be chosen within the outlined boundaries, $f \sim 50$ GHz an $d \sim 0.25$–1 m, to obtained the target order of magnitude of the density range and of the radial location of measurement. In practice, $f = 47$ GHz is convinient because of the availability of the hardware components for this frequency and $d = 42.5$ cm was chosen due to construction limitations [14]. The horn antennas sizes used are the same as for the reflectometer in [13], 20*27 mm$^2$ on the aperture and 2.39*4.78 mm$^2$ on the side connected to the waveguide. These values are used in further modelling.

## 3. Modelling setup

### 3.1 Modelling approach

Full-wave 3D simulations are performed in COMSOL commercial software, by using Finite Element Method in frequency domain. Among the built-in modules in COMSOL there is none suitable for simulating plasma similar to the tokamak edge conditions. Therefore, an approach developed specifically for such plasma is used, called RAPLICASOL [15]. Originally aimed at simulations in the domain of Ion Cyclotron Range of Frequency (ICRF), it has been widely benchmarked with other codes and experiments [16-20].

The RAPLICASOL approach consists of defining full-wave propagation inside a domain, which properties approximate cold plasma, and ensuring high single-pass absorption of the wave by a Perfectly Matched Layer (PML) on the outer boundary. It allows modelling a considerably smaller volume, rather than wave propagation and absorption at farther distance, and, as a result, such a model becomes computationslly feasible, while the physics inside the plasma domain can still be correctly reproduced without unwanted reflections from the outer boundaries.

Another important aspect of the RAPLICASOL approach is the detailed representation of the full antenna geometry, be it a whole ICRF antenna in other studies or a microwave horn launcher in this work. In the context of the current task, the wave interference taking place at the MILS receiver antenna and its transition to the waveguide can be accurately captured by using precise 3D geometry, and therefore the resulting phase and power measured at the end of the waveguide are obtained directly, without any additional processing of the signal mixing (as is necessary in ray-tracing, for example). This allows obtaining quantitative results of high accuracy, rather than qualitative description of the wave propagation and refraction. For the same reason, the simulations are carried out in 3D, even though a 2D model would be much simpler and could also give meaningful qualitative results.

Additional modelling for MILS is done using ray-tracing. This allows getting an insight into exact paths of different parts of the interferometer beam. Absolute values of the received phase and power have to be calculated from the contributions of all rays of the beam, which is a task much more challenging than the automatic calculation in the full wave simulations, and so the absolute calibration of the model is still work in progress. In this paper, only qualitative results of ray-tacing are demonstrated (Section 3.3).

### 3.2 Model components

In Fig. 1 the model geometry is depicted, with different materials shown in different colors. The coordinate system is chosen such that the axis of MILS is aligned to one of the axes ($y$-axis) and the two other axes are parallel to the edges of the horn antennas. The MILS axis is at $x = 0$, $z = 0$. This coordinate system does not coincide with the default coordinate system of ASDEX Upgrade, but it is much more convenient for the performed modelling (e.g. it allows simple and top-bottom symmetric density distribution definition).

COMSOL simulates inner volumes, filled with medium, while the volumes of the objects are "subtracted" and represented by boundary conditions. Therefore, the metallic horn antennas of MILS are represented by surfaces of Perfect Electric Conductor (PEC), which serve as the outer boundaries of the model, on one side, and border with the model volumes, on the other side. The existing symmetry in the toroidal direction is utilized to reduce the model size, as shown in the Fig. 1, only half of the whole model is simulated. On the symmetry plane $z = 0$, PEC material is defined to ensure correct field representation.

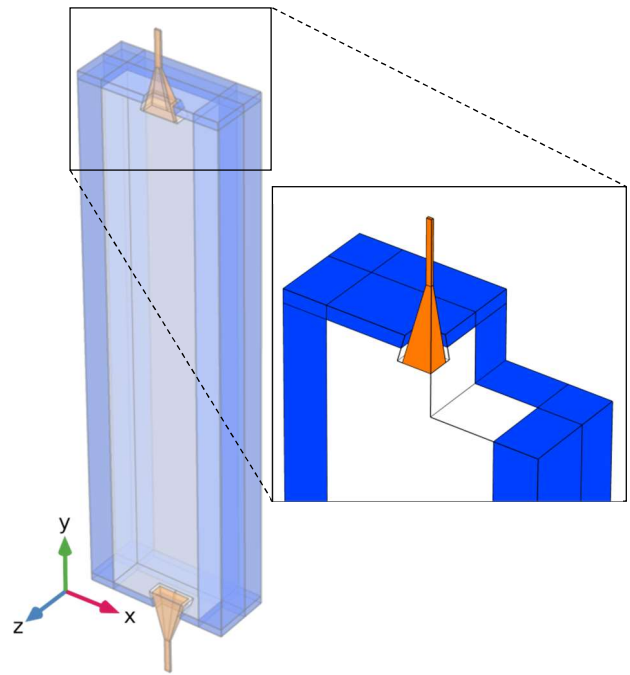

Fig. 1. Simulation geometry and materials: white - plasma, blue - PML, orange - vacuum. Full view with semi-transparency, zoomed-in part with a cut-out sector to demonstrate the inner structure.

The wave is excited on an active port at the end of the waveguide of the lower horn. The passive port at the upper waveguide detects the phase and amplitude of the incoming wave. There is no need to back it with a PML, because this kind of a port is able to absorb incoming waves, as long as they match the mode set for the port. Generally, not all scattered waves match the excited mode, but in our model this is guaranteed by the presence of the waveguide, which can support only a particular mode. To ensure that all other modes are damped before they can reach the port, it is crucial to place the port at a sufficient distance from the back of the horn antenna. In our model the waveguide length $l_{WG} = 3$ cm, which is more than three wavelengths $\lambda_{WG}$ of the wave at $f = 47$ GHz inside the used waveguide (cut-off frequency $f_{WG\ cut-off}$ = 31.4 GHz), as deduced from:

$$\lambda_{WG} = \frac{\lambda_{vac}}{\sqrt{\left(1-\left(\frac{\lambda}{\lambda_{WG\ cut-off}}\right)^2\right)}} = \frac{6.4\ mm}{\sqrt{\left(1-\left(\frac{6.4\ mm}{9.6\ mm}\right)^2\right)}} = 8.6\ mm \quad (1)$$

where $\lambda_{vac}$ is the wavelength in vacuum, in open space.

*3.3 Wave mode*

In a rectangular waveguide only modes, for which the cut-off frequency is less than the wave frequency, $f_c^{mn} < f$, can propagate. For both TE and TM modes:

$$f_c^{mn} = \frac{c}{2}\sqrt{\left(\frac{m}{a}\right)^2 + \left(\frac{n}{b}\right)^2} \quad (2)$$

where $c$ is the speed of light, $m$ and $n$ are mode numbers and $a$ and $b$ are waveguide sizes.

For WR19 waveguide ($a$ = 4.78 mm, $b$ = 2.39 mm):

$$f_c^{10} = \frac{c}{2a} = \frac{3e8}{2*4.78e-3} = 31.4\ GHz < f \quad (3)$$

$$f_c^{20} = \frac{c}{a} = 62.8\ GHz > f \quad (4)$$

$$f_c^{01} = \frac{c}{2b} = 62.8\ GHz > f \quad (5)$$

$$f_c^{11} = \frac{c}{2}\sqrt{\frac{1}{a^2} + \frac{1}{b^2}} = 70.2\ GHz > f \quad (6)$$

Therefore, the chosen frequency $f$ = 47 GHz provides a one-mode regime for the WR19 waveguide with TE10 mode.

The electric field is parallel to the short edge of the waveguide (and the horn antenna), therefore the horns are located with this edge aligned to the tokamak background magnetic field (z-axis in our models), to ensure O-mode propagation of the wave through plasma.

In homogeneous magnetized cold plasma, Maxwell's equations are solved for plane waves $\sim \exp(-i\omega t + i\mathbf{k} \cdot \mathbf{r})$ to obtain the wave equation [21]:

$$\mathbf{k} \times (\mathbf{k} \times \mathbf{E}) + \frac{\omega^2}{c^2}\bar{\bar{\varepsilon}}\mathbf{E} = 0, \quad (7)$$

with $\mathbf{k}$ being the wave vector, $\mathbf{r}$ the space vector and $\bar{\bar{\varepsilon}}$ the dielectric tensor defined as (for $\mathbf{B} \parallel \mathbf{z}$):

$$\bar{\bar{\varepsilon}} = \begin{pmatrix} \varepsilon_\perp & -i\varepsilon_X & 0 \\ i\varepsilon_X & \varepsilon_\perp & 0 \\ 0 & 0 & \varepsilon_\parallel \end{pmatrix} \quad (8)$$

with Stix parameters $\varepsilon_\parallel = 1 - \sum_s \frac{\omega_{ps}^2}{\omega^2}$, $\varepsilon_\perp = 1 - \sum_s \frac{\omega_{ps}^2}{(\omega^2 - \Omega_{cs}^2)}$ and $\varepsilon_X = \sum_s \frac{\Omega_{cs}\omega_{ps}^2}{\omega(\omega^2 - \Omega_{cs}^2)}$. $\Omega_{cs}$ is the Larmor frequency and $\omega_{ps}$ is the plasma frequency for plasma particles species s and $\omega = 2\pi f$ is the angular frequency of the wave.

At 47 GHz, the dielectric tensor components may be simplified to

$$\varepsilon_\parallel = \varepsilon_\perp = 1 - \frac{\omega_p^2}{\omega^2}, \varepsilon_X = 0 \quad (9)$$

Therefore, the O-mode dispersion relation ($\mathbf{k} \perp \mathbf{B}$, $\mathbf{E} \parallel \mathbf{B}$) for this frequency is:

$$N = \sqrt{1 - \frac{\omega_p^2}{\omega^2}} \quad (10)$$

where $N = \frac{kc}{\omega}$ is the refractive index.

The cut-off condition ($N$ = 0) is:

$$\omega = \omega_p = \sqrt{\frac{e^2 n_c}{\varepsilon_0 m_e}} \quad (11)$$

where $e$ is the electron charge, $m_e$ is the elctron mass, $\varepsilon_0$ is the vacuum permittivity and $n_c$ is the cut-off electron density, above which the wave cannot propagate. From eq. (8):

$$n_c = \frac{\varepsilon_0 m_e \omega^2}{e^2} \quad (12)$$

For 47 GHz, $n_c$ = 2.74*10[19] m[-3]. The wavelegth of the wave is infinite at $n = n_c$, and at lower densities it is $\lambda = \lambda_{vac}/\sqrt{1 - \frac{n}{n_c}}$.

When a wave passes through plasma, it experiences a phase shift, which, in the approximation of low densities ($\omega_p \ll \omega$, $\lambda \approx \lambda_{vac}$), can be written as:

$$\Delta\varphi = \frac{2\pi d}{\lambda_{vac}} - \frac{2\pi}{\lambda_{vac}}\int_0^d N(x)\,dx \approx \frac{2\pi}{\lambda_{vac}}\frac{n}{2n_c}d = \frac{\pi n d}{\lambda_{vac} n_c} \quad (13)$$

where $n$ is line-averaged electron density and $d$ is a distance over which the total shift of the wave phase was accumulated.

The exact integral cannot be calculated analytically, but a particular case of $n = const$ is useful for the calculations done in Section 5:

$$\Delta\varphi = \frac{2\pi d}{\lambda_{vac}} - \frac{2\pi}{\lambda_{vac}}\int_0^d N(x)\,dx = \frac{2\pi d}{\lambda_{vac}}\left(1 - \sqrt{1 - \frac{n}{n_c}}\right) \quad (14)$$

In conventional interferometer systems in plasma devices, the eq. (13) can be used to estimate line-average density and to localize the measured value to a certain central region of the plasma, where the highest density causes the strongest shift. This approach is inapplicable for MILS. The measured signal is not defined only by the part of the wave, which travels along the MILS axis. The detected phase and power change is the result of interference of the whole signal, which reaches the receiver horn (see Fig. 8). Several groups of rays from different plasma regions can contribute to the output at the same time (Fig. 2). The only way to interpret the measured data is to do a forward modelling of the diagnostic response to a certain plasma density.

*3.4 Definition of plasma and PML*

As shown on Fig. 1, the model uses three materials: vacuum, plasma and PML. In COMSOL a material can be defined with user-specified properties. By assigning the 3D dielectric tensor as in eq. (8), the permeability as $\mu$ = 1 and the electrical conductivity as $\sigma$ = 1, we define the cold magnetized plasma material. For our case of O-mode, the approximations in eq. (9) lead to a simpler case of $\varepsilon$ defined as a single value.

The PML material definition is similar to the plasma definition, but the permeability and permittivity tensors are different. The concept of PML functioning is based on transforming the incident propagating waves into evanescent ones. To achieve this, the model intrinsic real space coordinates are replaced within the PML domain by complex stretched coordinates:
$$x \to \int_0^x S_x(x')dx' \quad (15)$$
with $S_x(x) = 1 + \frac{i\sigma(x)}{\omega_0}$ as the stretching function. The damping function $\sigma(x)$ should be zero at the border of the PML, to ensure continuity, and be a rising function of the distance. The stretching functions used in RAPLICASOL are polynomial [15]:
$$S_x(x) = 1 + (S'_x + iS''_x)\left(\frac{x}{L_{PMLx}}\right)^{p_x}, x > 0 \quad (16)$$
with coefficients $p_x = 2$ being the order of the stretching function, $S'_x = 1$ the real stretch, $S''_x = -2$ the imaginary stretch and $L_{PMLx}$ the PML depth (6.5 mm in the direction of MILS axis and 21 mm perpendicular to it).

The dielectric tensor of the PML is given by:
$$\bar{\bar{\varepsilon}}_{PML} = \begin{pmatrix} \varepsilon_{xx}\frac{S_y(y)S_z(z)}{S_x(x)} & \varepsilon_{xy}S_z(z) & \varepsilon_{xz}S_y(y) \\ \varepsilon_{yx}S_z(z) & \varepsilon_{yy}\frac{S_z(z)S_x(x)}{S_y(y)} & \varepsilon_{yz}S_x(x) \\ \varepsilon_{zx}S_y(y) & \varepsilon_{zy}S_x(x) & \varepsilon_{zz}\frac{S_x(x)S_y(y)}{S_z(z)} \end{pmatrix} \quad (17)$$
And the permeability tensor is definied analogously.

*3.5 Model output*

The model output is the power and phase deviation in plasma compared to vacuum. To obtain these values, a model filled with vacuum instead of plasma was simulated as a reference. The values presented in the paper are $\Delta\varphi = \varphi_{vac} - \varphi_{pl}$ and $\frac{P_{pl}}{P_{vac}}$.

The power is calculated in COMSOL from the $S_{21}$ parameter of the 2-port model. $S_{21} = \sqrt{\frac{P_2}{P_1}}$, where $P_1$ is the power specified on the port of the emitter horn antenna and $P_2$ is the power detected at the receiver port. Since $P_1$ is the same for both vacuum and plasma cases, $\frac{P}{P_{vac}} = \left(\frac{S_{21\,pl}}{S_{21\,vac}}\right)^2$.

The phase is obtained as $\varphi = \arg(E_z)$, where $E_z$ is the only non-zero component of the electric field for TE10 mode. The evaluation is done at a point at the center of the receiver waveguide. It is useful to note here that in COMSOL the definition of this point has to be done globally, and not through the geometrical elements, otherwise unexplained variations have been noticed, which lead to a wrong result.

## 4. Density profile

*4.1 Shape of the radial density profile*

The temporal resolution of a microwave interferometry can be extremly high, and in practice it is most often defined by the available data acquisition system rather than by the diagnostic limitations. Therefore, the variations in the plasma density can be observed with the necessary precision in time with MILS diagnostic and a study of such variations can be carried out. However, in this paper transient SOL events, such as Edge Localized Modes (ELMs), are not considered, and the data interpretation is limited to a monotonic (inter-ELM-like) density profile shape. In experiments, possible microturbulence along the wave propagation path get averaged, as typical for an interferometer, so the reconstructed density profile should be considered as averaged over such microstructures.

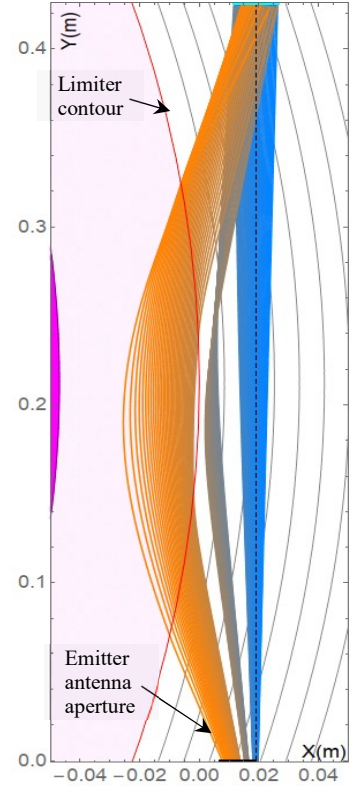

Fig. 2. Qualitative picture of microwave beam rays propagating in and being refracted by plasma. Several groups of rays, emitted at different angles and passing through different plasma layers, reach the receiver and contribute to the measured values.

In order to build a universal database for density evaluation from the experimental data of MILS, our approach is to fix a certain shape of the 1D radial density profile. The main limitation on the precision of the reconstruction of the density porfile shape, from the point of view of the interpretative modelling, is the unavoidable simplification of the shape. The 2D space $(\varphi, P)$ of measurements needs to be mapped into the modelling parameter space, which describes the shape of the profile. The larger the number of parameters is, the more computationally expensive it becomes to build a database.

One of the simplest possible profile shapes, which should still describe the actual density relatively well, is the one shown in Fig. 3, as a function of the radial distance to the limiter $R - R_{lim}$. The inner and outer density layer (relative to the limiter contour) have constant exponential decay with coefficients $a_{in}$ and $a_{out}$ respectively, with the density considered to typically decay faster in the outer part (limiter shadow), where the connection length of the magnetic field lines decreases greatly [22], so the absolute value of $a_{out}$ is always bigger than that of $a_{in}$. In reality, the transition in the decay length between the inner and outer density regions should be smoother, since the connection length also does not go sharply from the regime of tens of meters long field lines reaching the divertor to ~ 1 m value corresponding to the toroidal distance between the limiters. The plasma shape fit to the wall is never perfect, so there is always an intermediate region where the magnetic field lines intersect the first wall sections, baffles, etc.

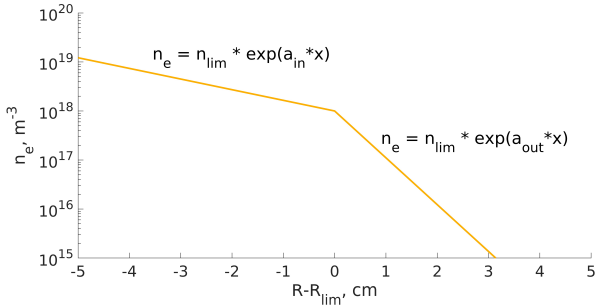

Fig. 3. Shape of the radial density profile used for modelling, made of two exponential functions, with a transition at the limiter.

Therefore, the considered profile shape is a simplification, adopted for the current analysis in order to enable a parametric description of the modelled density. Such profile is defined by three parameters: exponential coefficients $a_{in}$, $a_{out}$ and $n_{lim}$ (density at the limiter position, where the profile bends). The task of the modelling analysis is, therefore, to provide a mapping from $(n_{lim}, a_{in}, a_{out})$ to $(\varphi, P)$.

*4.2 3D density definition*

In the previous subsection a 1D density profile was discussed, limited to the radial direction. In the experimental conditions, various reasons can cause poloidal (filaments, turbulence, etc.) or toroidal (magnetic perturbations, local gas puff, etc.) density inhomogeneity. It is out of the scope of this study to consider them, so the 3D density in the presented models employ density homogeneous along the poloidal and toroidal coordinate.

In order to excite the O-mode wave, the axis of MILS is directed perpendicularly to the background magetic field. In the radial-toroidal cross-section the position of the horn antennas is chosen such as to ensure symmetric tangential crossing of the plasma. The limiter contour region near MILS can be approximated by a circle and the flux surfaces by circles concentric to it. The distance between the limiter and the MILS axis in its center is $d_{axis-lim} = 1.91$ cm. The resulting 2D shape of the density used in the current modelling is shown in Fig. 4. Constant colour corresponds to constant density and the shape is the same in all models, irrespective to absolute $n_e$ values.

The 3D density profile is obtained by defining $n_e = const$ in the direction perpendicular to the plane shown in Fig. 4, i.e. toroidally. Unlike the complex and changing conditions of the signal collection in the poloidal-radial plane, along the toroidal direction it is always a very narrow region approximately of the size of the horn in this direction (2.7 cm), where the wave detection takes place. Therefore, the influence of any toroidal variations can be considered as extremely small.

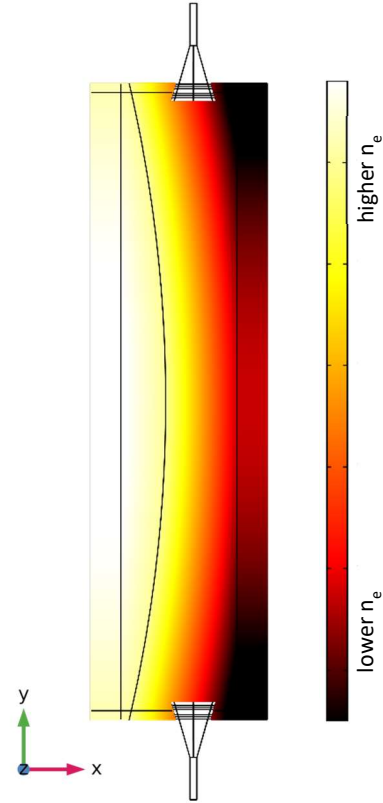

Fig. 4. Shape of the density profile on a 2D slice in the radial-poloidal plane. Density is constant on the circles concentric to the circle that approximates the limiter (black curve). Vacuum is used inside the horn antennas.

In the experiment, horn antennas are immersed into plasma and there might be non-zero density inside the antennas themselves. However, this value would be very low and its influence on such short distance cannot play any significant role on the signal detection. In the models, horns volume is therefore filled with vacuum.

The described 3D density distribution definition is used in the modelling of the interferometer wave propagation through the plasma.

**5. Basic tests for model validation**

The model can be verified on simple cases where the influence of the plasma density distribution on the wave propagation can be calculated analytically. A test was carried out with plasma of constant density of various magnitudes. In Fig. 5 the simulation results are compared to the theoretical prediction given by eq. (14). For the theoretical curve, the divergence between the exact theoretical formula and its linear approximation is not strong, but noticeable at the higher density. To compare exactly the theory and simulation, it is important to take this into account. The deviation of modelling results is shown in Fig. 6. At low densities the simulation gives slightly lower values, while it overestimates the phase shift at higher densities, relative to the theoretical prediction.

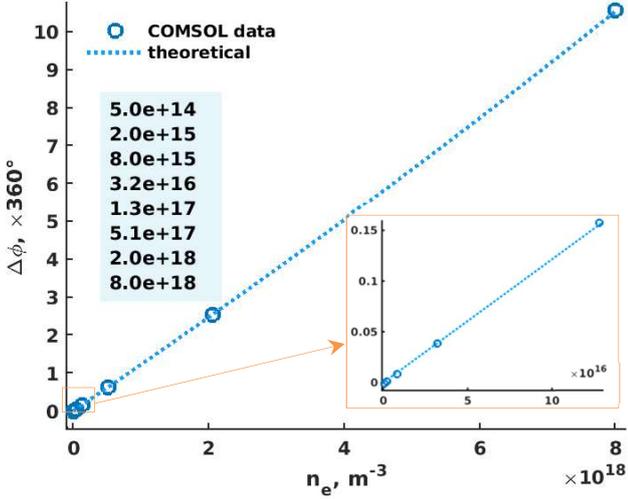

Fig. 5. Comparison of simulated and theoretical phase shift of the wave in plasma of constant density, as a fraction of full 360° turns. A zoom-in on the low densities is shown in a box.

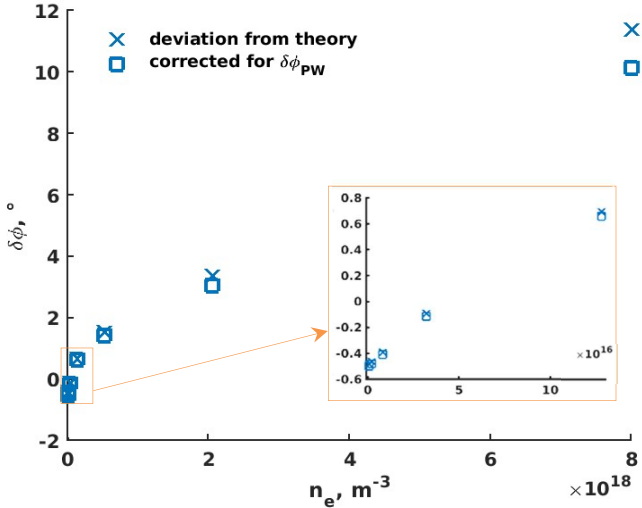

Fig. 6. Difference between the modelling and theory as a function of density, for the data in Fig. 5. Additionally, values corrected for the numerical error obtained from plane wave simulations $\delta\varphi_{PW}$ are plotted. A zoom-in on the low densities is shown in a box.

We have identified two reasons, which can contribute to the observed deviation. The first one is the numerical precision of the solution. It was assessed on a model without horn antennas, but with the same model parameters. A plane wave, excited at one side of a 3D rectangular parallelepiped, reaches a receiver port at the opposite side and the phase difference between the source and the receiver is proportional to the distance. A scan with the same density values set was performed. The obtained deviation from analytical calculation (distance divided by wavelength), relative to vacuum, is density dependent and can be well fitted by a linear function $\delta\varphi_{PW} = 1.533 * 10^{-19} * n_e - 0.017$ [°] with root-mean-square error of 0.027°. The resulting $\Delta\varphi$ deviation from theory, when the error from the plane wave model is taken into account, is shown in Fig. 6. Its influence is small compared to the absolute values of the deviation.

Another possible cause of the phase deviation from theory is that the horn antennas and waveguides convert the incident wave, present at the horn aperture, to the waveguide fundamental mode, of which the phase gets measured at the end of the waveguide, in a complex 3D way. Part of the incident wave is reflected in the process (see Fig. 8 and Fig. 9). Therefore, the measured phase shift in a model with constant density plasma is defined not only by the layer of plasma between the horns, but also by the conversion of the signal by the horn itself. This effect can be hardly characterized numerically, but it can be seen that its amplitude is density dependent (Fig. 6).

In further cases of plasma density profiles more complex than constant, the same 3D effect of the horn influence on the detected phase and power should be present. As the same effect can be expected in the experiment, the advantage of the full-wave simulation, when compared to ray-tracing, is that such effect can be realistically reproduced and taken into account in the modelling.

## 6. Simulations with density profiles

### 6.1 Optimized model

Each density profile requires a model run. The chosen radial density profile shape can be described by three parameters, which means that, in order to explore the 3D parameter space, the number of simulation runs to be done is ≈ (number of points for each parameter)³. To enable large parameter scans, the model has to be optimized as much as possible, to reduce the memory usage and computation time. Following steps were taken during the optimization:

- Quadratic discretization was compared to and replaced by the cubic discretization (increasing the discretization order approximately corresponds to a uniform mesh refinement);
- Mesh was scaled in the direction perpendicular to the MILS axis, relative to the direction along the axis, to reduce the number of mesh elements;
- Mesh element size was scanned to find the largest acceptable value;

- Plasma domain sizes were varied to study their influence on the model output and the minimal possible size was used.

The details of this optimization are given in the Appendix. All the results presented in the paper have been obtained from the model with the same optimized settings: discretization, mesh shape, size and scale. The only varied characteristic was the size of the model in $x$-direction. Two sizes were used, a smaller model was suitable for the cases when the wave does not penetrate deep into the plasma (higher density cases) and a bigger model was needed for lower densities, to ensure that the whole region where the wave propagates is simulated. The details are described in Appendix, but the key point is that the results from both bigger and smaller models can be analyzed together as one set of data. The errorbars in the final model are quite small, which is fortunate, considering the fact that the model had to be optimized for the performance. The systematic errors are $\delta_{\varphi\,sys} = -1°$ and $\delta_{P\,sys} = +1\%$ and the random errors are $\delta_{\varphi\,rnd} < 0.5°$ and $\delta_{P\,rnd} = 1.5\%$.

*6.2 Results*

The mapping from the parameter space $(a_{in}, a_{out}, n_{lim})$ to the measured parameters $(\Delta\varphi, P/P_{vac})$ is presented in Fig. 7. The parameters, which define the density profile, correspond to the axes. Larger absolute values of $a_{in}$ and $a_{out}$ mean more strongly decaying density profile, while a shift along the $n_{lim}$ axis leads to an overall increase/decrease of the density in the whole profile. $\Delta\varphi$ is shown as the color of the markers and $P/P_{vac}$ is indicated with marker size, bigger for larger values.

While it might be difficult to grasp all the details from the 2D view of this complex plot, some important aspects can be noticed. There are clear dependences:

- $\Delta\varphi$ increases with increasing density, as expected from the general interferometer theory;
- prominent variation of $P/P_{vac}$ depending on the density is present;
- however, the variation of $\Delta\varphi$ and $P/P_{vac}$ is clearly more complex as it could have been, if it was influenced only by the density along the MILS axis; the interference of several parts of the beam gives the resulting output (Fig. 8);
- the rise of $\Delta\varphi$ from low to high values is monotonic for nearly the whole region, with two exceptions: pronouncedly non-monotonic around $n_{lim} = 1*10^{16}$–$5*10^{17}$ m$^{-3}$; and weakly non-monotonic around $n_{lim} = 1*10^{18}$–$2*10^{18}$ m$^{-3}$ when $a_{in} > -50$;
- the power of the received signal lowers monotonically, starting from $n_{lim} \sim 0.5$–$1*10^{18}$ m$^{-3}$, with increasing density, due to the wave being sronger refracted by higher density plasma;
- the variation of $P/P_{vac}$ in the region of lower densities is more complex, as it is caused by the competing processes of wave partial refraction out of the receiver and at the same time wave "focusing" from the high-density region where it would miss the receiver in the vacuum case; the wave focusing reaches its maximum in the region $n_{lim} \sim 0.5$–$1*10^{18}$ m$^{-3}$ and is illustrated in Fig. 9, and can be qualitatively observed in Fig. 2;
- in the region of lower densities, an interesting effect is observed – negative phase change; while in other regions the phase shift is moslty caused by the change in the refractive index, in this part an additional factor starts to play a role – the longer path of the wave propagation compared to the path along the MILS axis;
- there are regions where both $\Delta\varphi$ and $P/P_{vac}$ do not vary significantly along one of the axes, thereby the parameter space degenerates from 3D to 2D: (nearly) no dependence on $a_{in}$ for $n_{lim} > \sim 4*10^{18}$ m$^{-3}$ and (nearly) no dependence on $a_{out}$ for $n_{lim} < \sim 3*10^{17}$ m$^{-3}$; the reason is that not the whole simulated density profile influences the measured signal (more details in Section 7);
- a more general conclusion than in the previous point is that the influence on the output $\Delta\varphi$ and $P/P_{vac}$ from different groups of rays (Fig. 2) has different magnitude; from a radial density profile, its central part has the largest influence, while the values of density on the profile edges only weakly affect $\Delta\varphi$ and $P/P_{vac}$ (it will be also seen in Section 8);
- not the whole 3D parameter space within given boundaries needs to be scanned, since some values combinations would result in too unrealistic density profiles (e.g. too steep density increase towards the plasma core for large $n_{lim}$, if too large abs($a_{in}$) is taken, etc);
- gradients are different in different regions and depend on the direction, therefore some regions can be investigated with fewer points and others need fine resolution in the parameter space, which explains the varied density of modelling points in Fig. 7.

The results of simulations can be also represented as a relation between the two measured quantities: $P/P_{vac} = f(\Delta\varphi)$, see Fig. 10. There is a peculiar pattern formed by the combinations of the two values and it can be seen that not any combination is possible, but only a certain range. To connect this figure to the plot in Fig. 7, the markers are shown with colors indicating the $n_{lim}$ value. It can be seen in which direction $P/P_{vac}$ and $\Delta\varphi$ generally shift with growing density.

**7. Measurement region determination**

During the database construction, it was observed that the region where the beam part, which reaches the receiver, propagates varies considerably for different plasma densities. A study was performed to identify the part of the radial density profile, which makes significant influence on the measured quantities.

In all models, the initial limits are defined by the plasma domain sizes (see Appendix for more details on sizes),

therefore the range is $R - R_{lim}$ = (-3;5) cm or (-5;5) cm, for higher and lower density range, respectively. At the low-$n_e$ side, a simple criterion was chosen to cut the density. $n_e$ = 2.3*10$^{15}$ m$^{-3}$ corresponds to 1° phase shift over the MILS axis length of 42.5 cm. This value is taken as the lowest limit and everything below it is cut from the reconstructed density profile.

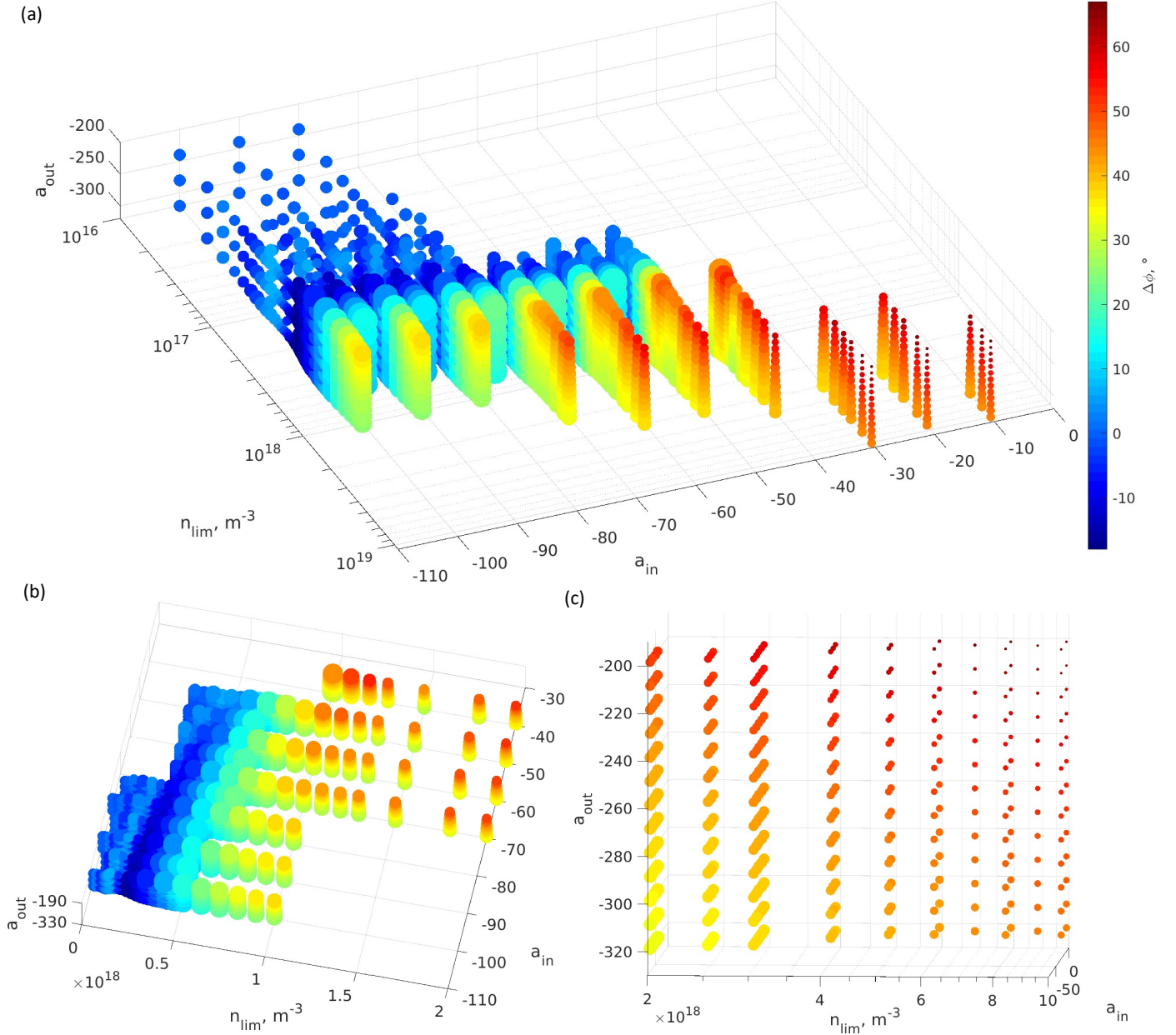

Fig. 7. (a) Measured values $\Delta\varphi$, $P/P_{vac}$ as functions of 3 density profile shape parameters: $n_{lim}, a_{in}, a_{out}$. $\Delta\varphi$ is shown by marker color and $P/P_{vac}$ by marker size. (b) Zoom on low densities, with linear scale of $n_{lim}$ axis. (c) Zoom on high densities. An interactive version of this figure is accessible online.

In order to find out to which part of the profile at the high-$n_e$ side the measured vales are not sensitive, the profile was partially cut, more at each step, and replaced by vacuum. The step of the parameter scan was set to 0.5 cm. The models with trimmed density were compared to the original models and then the largest possible cut was chosen, such that the $\Delta\varphi$ and $P/P_{vac}$ deviations were within the corresponding random errors, obtained during the model optimization (see Appendix), 0.5° and 1.5 %.

The result is shown in Fig. 11, on a 3D plot with the $a_{out}$ axis being perpendicular to the plot plane. Since the density is cut in the region where it is defined by the $a_{in}$ parameter, it is not expected to have a large dependence on $a_{out}$. However, for all $(n_{lim}, a_{in})$ combinations it was checked by simulating 2 points along $a_{out}$, for its minimum and maximum. For some points a dependence is observed, and they are plotted as 2 markers of different colors at one location, the smaller one corresponding to the minimum $|a_{out}|$.

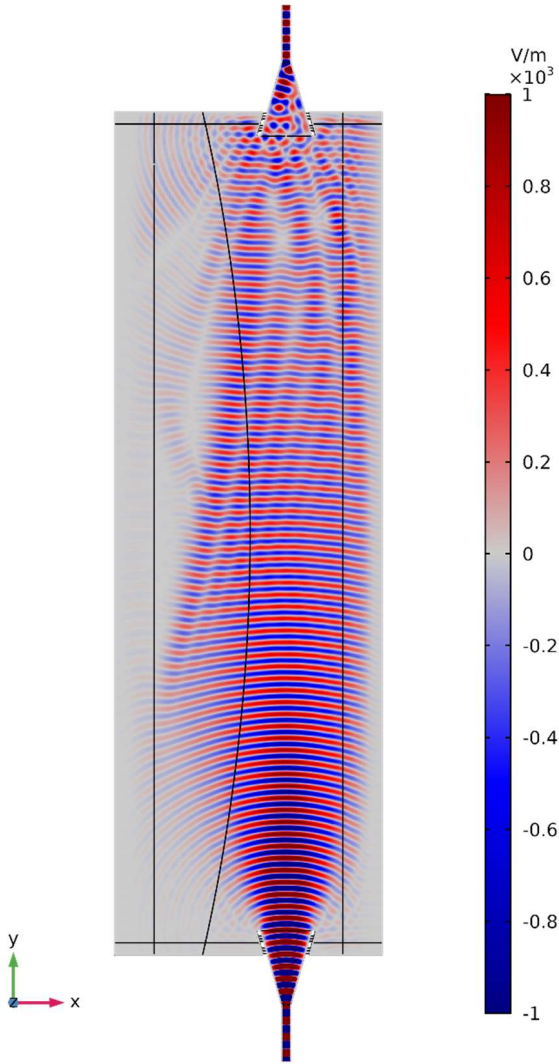

Fig. 8. Distribution of MILS electric field (main component $E_z$) in plasma with radial density profile (0.3e18,-80,-240). Beam parts propagate through different layers of plasma and their interference in the horn antenna leads to the measured output.

Based on the obtained results, limits were set for the density at each $R - R_{lim}$ point, above which the point should be cut from the density profile (dashed lines):
- $n_e > 6.5 * 10^{18}$ m$^{-3}$ at $R - R_{lim} = -3.5$ cm
- $n_e > 5.0 * 10^{18}$ m$^{-3}$ at $R - R_{lim} = -3$ cm
- $n_e > 4.2 * 10^{18}$ m$^{-3}$ at $R - R_{lim} = -2.5$ cm
- $n_e > 4.2 * 10^{18}$ m$^{-3}$ at $R - R_{lim} = -2$ cm
- $n_e > 4.2 * 10^{18}$ m$^{-3}$ at $R - R_{lim} = -1.5$ cm
- $n_e > 4.2 * 10^{18}$ m$^{-3}$ at $R - R_{lim} = -1$ cm
- $n_e > 3.6 * 10^{18}$ m$^{-3}$ at $R - R_{lim} = -0.5$ cm

It can be seen that these lines are not exactly at the borders of marker clusters for each color. They are chosen such that no significants radial points are cut out, and rather an extra point might be left uncut in some profiles. Such radial points are in any case relevant, just with lower precision, since the radial points are not independent from each other.

The part of $(n_{lim}, a_{in})$ parameter space, shown in gray in Fig. 11, with $n_{lim} \geq 4*10^{18}$ m$^{-3}$, has a separate definition of limits, because of too different $r_{cut}(n_{lim}, a_{in})$ behaviour:
- $n_e > 5.0 * 10^{18}$ m$^{-3}$ at $R - R_{lim} = -3$ cm $-->$ limit at -3 cm
- $n_e > 7.9 * 10^{18}$ m$^{-3}$ at $R - R_{lim} = -2.5$ cm $-->$ limit at -0.5 cm
- $n_e > 7.9 * 10^{18}$ m$^{-3}$ at $R - R_{lim} = 0$ cm $-->$ limit at 0 cm

This additional definition of limits is rather complicated, but unfortunately a simple uniform condition cannot be constructed.

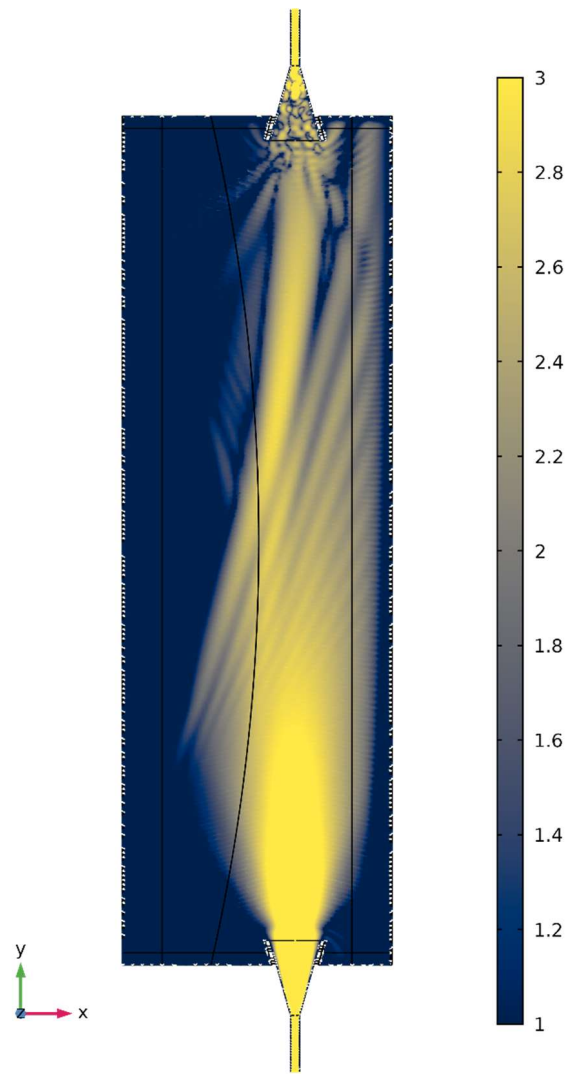

Fig. 9. Wave focusing to the receiver: part of the wave on the left, which in vacuum cannot reach the receiver, is refracted by the plasma and gets detected. Poynting vector main component $S_y$ in log10 scale is plotted. Density profile (0.9e18,-60,-280).

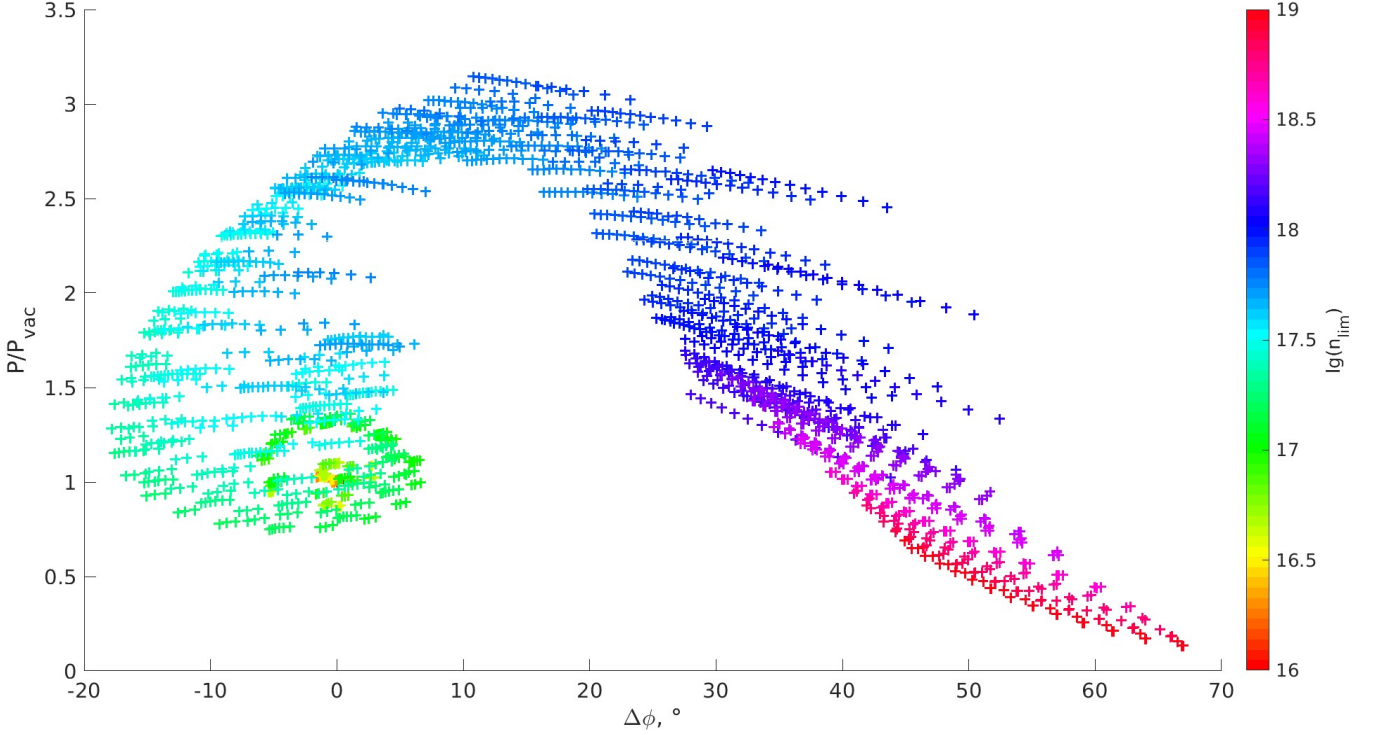

Fig. 10. Interdependence of phase and power measured in the simulations. Color corresponds to the value of $n_{lim}$ for each point.

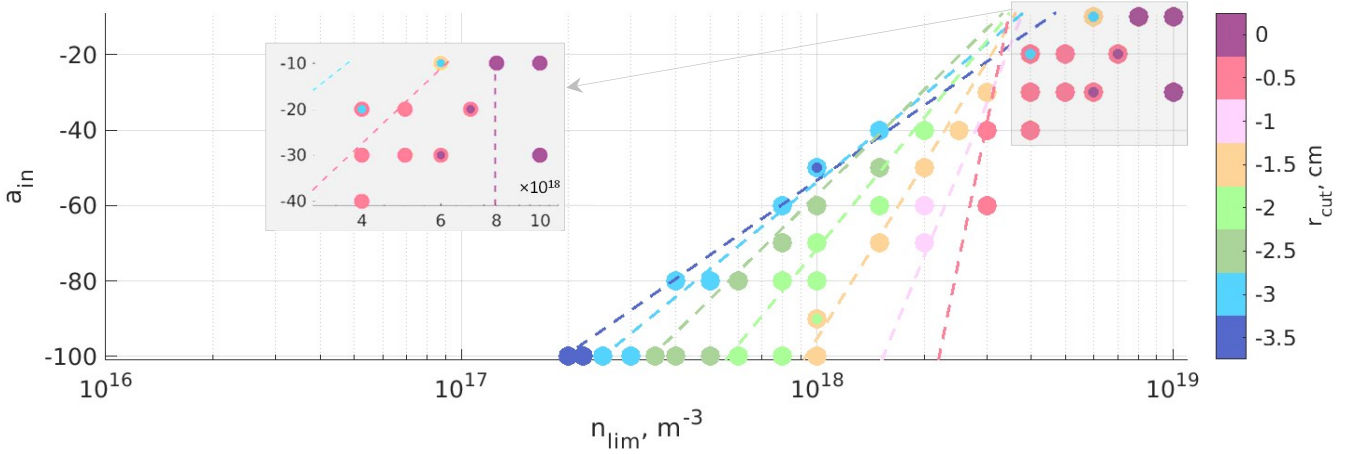

Fig. 11. Distribution of the largest possible cut position as a function of density profile parameters $(n_{lim}, a_{in})$. For each point two values of $a_{aout}$ are taken and the poins with lower $|a_{aout}|$ are plotted with smaller size, so that both points can be seen in the case of defferences along $a_{aout}$. Dashed lines show the chosen limits for the density radial cut, with colors corresponding to $r_{cut}$.

An overview of limitations on density and radial range is given in Fig. 12. The database, shown in Fig. 7, is used to plot the density profiles defined by the set of parameters $(n_{lim}, a_{in}, a_{out})$ and the radial cuts are applied according to the established procedure.

## 8. Density profile reconstruction

For the reconstruction of a density profile from the output of MILS, we establish a dedicated procedure named genMILS. The procedure is based on a genetic algorithm [23] of model selection. As a problem-solving strategy, genetic algorithms have been successfully applied to a wide variaty of tasks in various fields of science.

In genMILS, points of the constructed database are considered as possible candidates and the optimization procedure consists of finding the best combination of the candidate models for an input of $\Delta\varphi$ and $P/P_{vac}$. For an input point, all surrounding points in the range $(3\delta_\varphi, 3\delta_P)$ are selected, where $\delta_\varphi = 1°$, $\delta_P = 3\%$ are errorbars for $\Delta\varphi$ and $P/P_{vac}$ respectively, defined for this study. This level of errors could be expected in the experimental measurements and the values are twice larger than the random errors, calculated for the numerical model of MILS syntetic diagnostic (see Appendix). The systematic errors of numerical calulations are taken into account by shifting the phase-power diagram correspondingly (see Appendix).

Each of the surrounding points gets assigned an Akaike weight $w$ [24] depending on the distance to the input point on the phase-power diagram, under the assumption of Gaussian distribution of $P$ and $\varphi$ errors:

$$w = \exp\left(-\frac{(\varphi-\varphi_{input})^2}{2\delta_\varphi^2} - \frac{\left(\frac{P-P_{input}}{P_{input}}\right)^2}{2\delta_P^2}\right) \quad (18)$$

In order to calculate the density from the weighed contributions, we describe the density profile not as a whole, but distribute radial points along it. The set of radial points in the density profile constitutes the set of genes in our genetic algorithm. The resulting $n_e$ at each radial point is

$$n_e = \frac{\sum_i n_{e_i} * w_i}{\sum_i w_i} \quad (19)$$

where $n_{e_i}$ is the density value at this radial position in a profile number $i$ from the contributing profiles.

Some radial points (or segments) of the density profile contribute dominantly to the measured signal, and some influence it only weakly. The genetic algorithm allows us to identify the most important genes automatically for each case. Profiles with similar dominant genes lie close to each other on the phase-power diagram, but they can differ noticeably in some less crucial radial points. A profile reconstruction from such neigbours should take into account different level of confidence for different radial points, depending on whether they represent dominant or minor genes. It is implemented by assigning an error at each radial point. The root mean square error of $n_e$ is calculated as

$$\delta_{n_e} = \sqrt{\frac{\sum_i (n_e - n_{e_i})^2 * w_i^2}{\sum_i w_i^2}} \quad (20)$$

Each radial point is processed separately, as if they were independent, but one has to remember that they are not independent, as well as their errors.

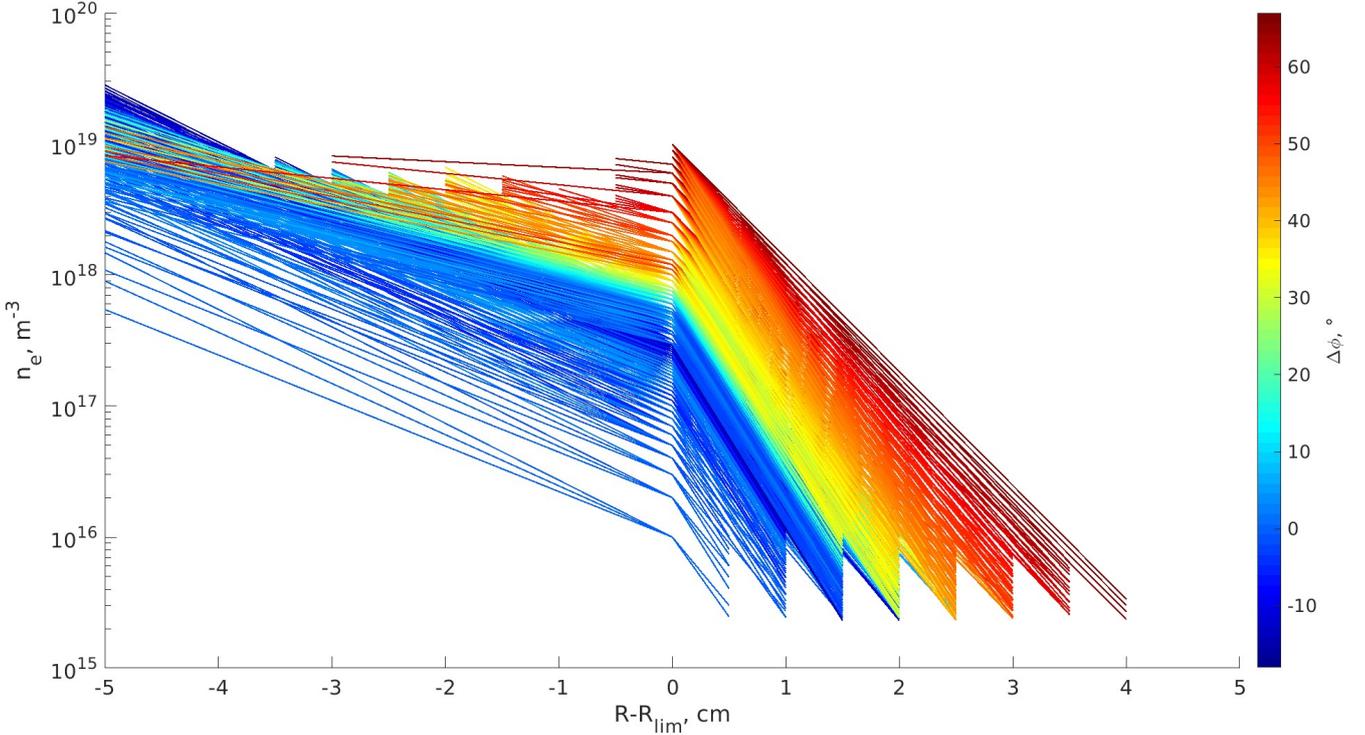

Fig. 12. Illustration of the measurement range limitations, both for density values and for radial locations. Each profile corresponds to a point in Fig. 7, with $\Delta\varphi$ shown by line color. Note that many profiles partially overlap and not all of them can be fully seen.

In order to demonstrate the capability of density reconstruction using the developed genMILS method and to estimate the errorbars of the obtained density values, several examples are taken, distributed along the phase-power diagram. The points, taken as example input, and their chosen neighbours are plotted in Fig. 13, with weights shown in color.

Various cases are taken, with less or more neighbours, homogeneously distributed on all sides or present mostly on one side, with less or more points in close proximity to the input point. The resulting density profiles are shown in Fig. 14. A step of 0.5 cm was chosen between the points along the radial coordinate.

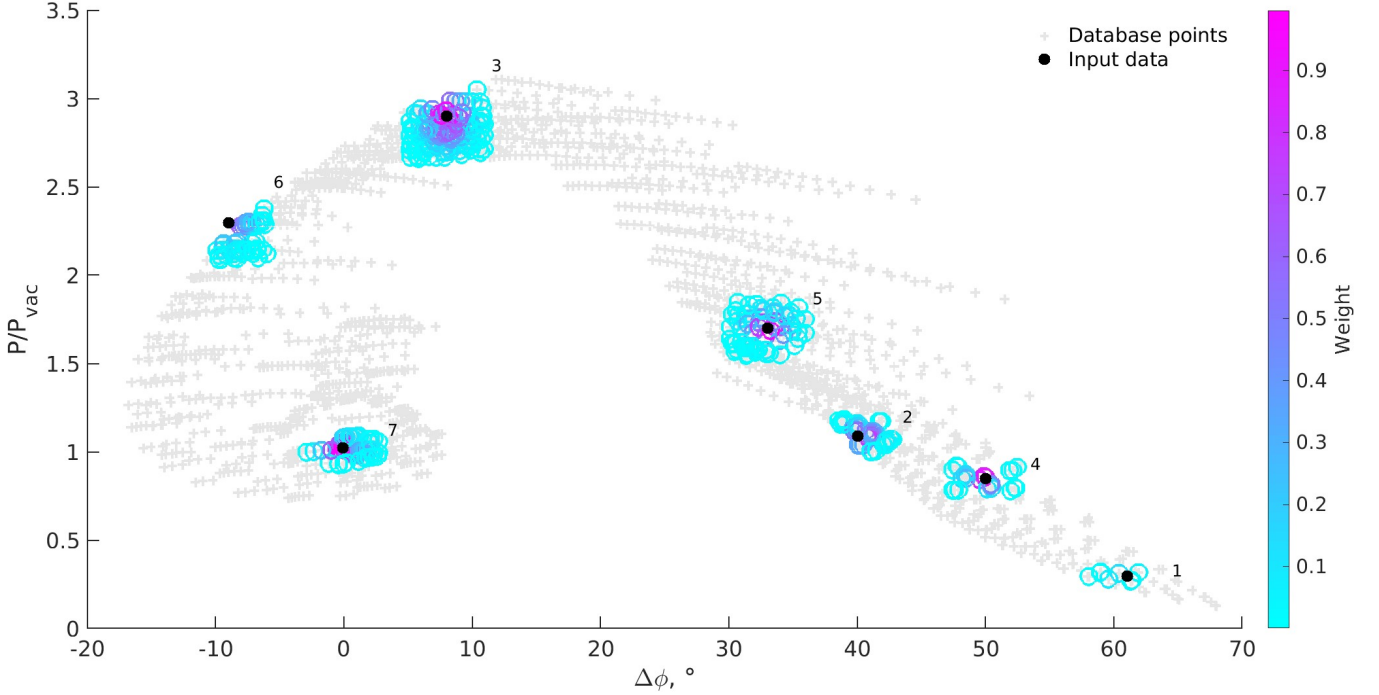

Fig. 13. Points chosen as examples for density reconsrtuction. Around each black dot marker, corresponding to the input data, a set of closest poins is shown with circle markers, their colors corresponding to their weight in the density profile calculatuion.

The resulting errrors are summarized in Table 1. Unless going to the smallest densities, or considering the points at the very edge of the radial profiles, the errors of 5–15 % can be expected for the reconstructed density. At densities $10^{16}$–$10^{17}$ m$^{-3}$ the errors might exceed 15 % in some cases. At densities $\leq 10^{16}$ m$^{-3}$ the order of value can be reliably predicted, but more precise estimation is challenging. In the part of the phase-power diagram close to (0,1), i.e. vacuum, the prediction becomes complicated, due to non-monotonic and large variations of $\Delta\varphi$ and $P/P_{vac}$ between neighbouring points in the parameter space (Fig. 7). Good accuracy is achieved for the range of densities relevant to the ASDEX Upgrade edge, which corresponds to the space on the phase-power diagram approximately from the example profile #6 and further towards higher densities (Fig. 13). This part is distributed over a large space (compared to the low-density part), so even quite similar profiles can be distinguished from each other, which leads to high accuracy in density prediction.

The developed genetic algorithm, which uses weighed contributions of the database points found in the vicinity of the input value, provides a robust way to choose more correct density profile between several options, which have differences in some radial parts larger that the errorbars estimated above. It is illustrated by the example of density profile reconstruciton #2. One of the neigbour points closest to the input has a significant weight and it differs noticeably from other closest neighbours (Fig. 14a, thick dashed lined, lowest for #2 in the region from -1 to 1 cm), with, for example, $n(R - R_{lim} = 0)$ twice lower than the reconstructed value at this radial point and with higher densities at $R - R_{lim} > 1$ cm compared to other neighbour profiles. Even though this point from the database has $\Delta\varphi$ and $P/P_{vac}$ close to the rest of chosen neigbours, the wave propagation and refraction for this case of plasma density profile is quite different from them. Different groups of rays make the largest contribution to the output values, in this case it can be seen that denser plasma in the limiter shadow could lead to the same power and phase output for this particular profile, but other profiles with similar density around $R - R_{lim} \leq 1$ cm already have too different values of $\Delta\varphi$ and $P/P_{vac}$ and therefore are not selected to the list of the input point neighbours. They do not possess the needed distinct features, in other words the dominant genes, which would allow them to be selected in our genetic algorithm. The distinct features for the neighbours of the input point #2 turns out to be the part of the density profile around approximately $R - R_{lim}$ = 0 to 1.5 cm. As illustrated with this example, the whole diagram of $\Delta\varphi$ and $P/P_{vac}$ (Fig. 10) consists not of randomly dispersed combinations of phase and power, but of points, grouped by certain distinct features.

| # | mean $\delta_{n_e}$, % | min $\delta_{n_e}$, % | max $\delta_{n_e}$, % |
|---|---|---|---|
| 1 | 7.2 | 3.2 | 12.1 |
| 2 | 19.9 | 4.5 | 47.0 |
| 3 | 22.1 | 3.7 | 61.5 |
| 4 | 6.9 | 2.9 | 11.7 |
| 5 | 16.5 | 4.3 | 32.8 |
| 6 | 10.4 | 1.5 | 26.0 |
| 7 | 76.6 | 56.3 | 128.8 |

Table 1. Mean, minimun and maximum value of errors for each of 7 reconstructed density profiles.

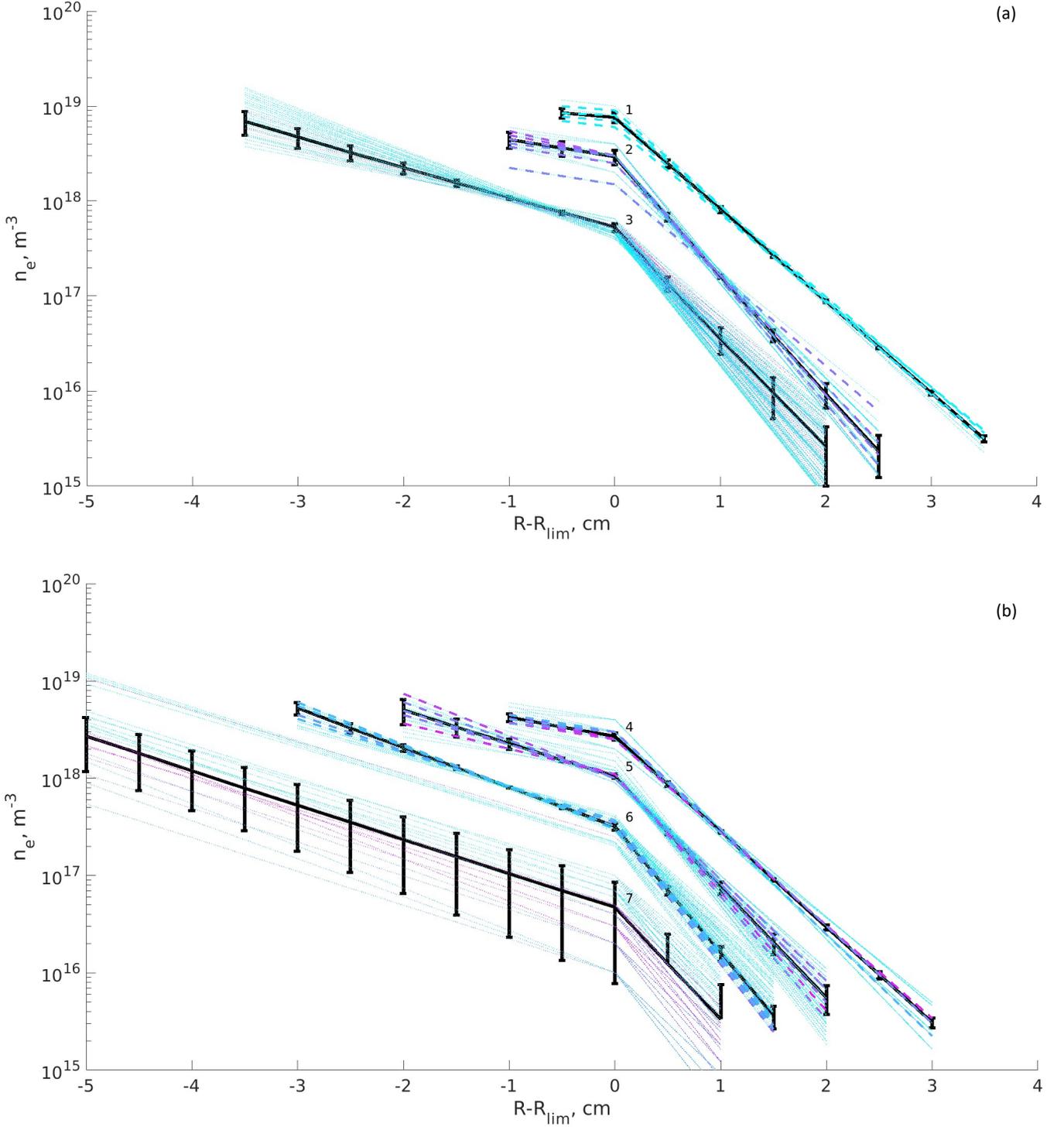

Fig. 14. (a), (b) A set of examples of density reconstruction. The solid lines with errorbars show the final profiles, the dashed thick lines display the profiles, used for fitting, which had the biggest weight (> 5% of the total), and dotted thin lines – profiles with weight ≤ 5% of the total. Numbers near the profiles correspond to numbers in Fig. 12. Colors correspond to the weight, as in the colorbar of Fig. 12.

Consequently, the independent evaluation of density at several radial points, which are not defined as independent in the parametrized profile shape, is justified by the observed grouping of the points by their genetic similarities. The reason for it lies in the specifics of the wave propagation through tokamak-edge-like, curved plasma. Each radial bin of a density profile brings a certain percentage of contribution to the resulting values of $\Delta\varphi$ and $P/P_{vac}$ and this percentage can differ significantly between radial bins of a profile. A reconstructed density profile, in general, does not have to take the shape of the analytical profile defined for database points. It still resembles it rather closely, because the shape is defined by just three parameters. In that sense, a possible way for improvement can be connected with increasing genetic variety of the database models. It has to be done in a thoughtful way, otherwise it might turn out as too computationally demanding and would at some point bring us to the limit of model complexity, which can be resolved by means of two measurable quantities.

## 9. Discussion

### 9.1 The presented analysis

The key conclusion of the performed analysis is the proof of capability of the newly developed diagnostic technique MILS, in combination with the presented data interpretation algorithm genMILS, to provide radial density profile reconstruction along several cm of the edge plasma. The physical principle of the wave refraction and phase shift in plasma, utilized for MILS, results in the two measured values, $\Delta\varphi$ and $P/P_{vac}$, forming a specific pattern of possible combinations, where points are grouped by their distinct features (profiles with similar density parts, which make the largest contribution to the output, are close on the phase-power diagram). By using the developed genetic algorithm for density reconstruction, which takes into account different contributions of different radial points, the errors are kept sufficiently low for a large range of plasma densities, with the assumption that the chosen shape of the density profile is able to capture the main distinct features of experimental density profile, which determine MILS wave refraction and phase shift in the region of measurement.

As was mentioned earlier, the main limitation of the presented approach is the radial profile shape description by 3 parameters. The number of parameters needs to be increased in order to reproduce the experimental radial density profile in more details. The performed modelling provides experience to facilitate a construction of an expanded database, so the increase of the number of parameters to 4–5 is feasible and some limited sets of data could be also analysed with higher parameter numbers.

### 9.2 Possible diagnostic improvements

A significant improvement of the precision of density reconstruction can be achived by using more than one receiver antenna for MILS. Even one more additional receiver can lead to doubled number of the measured quantities, so the mapping from those to the parameter space describing the density profile would become substantially more precise.

More information about the density can be also obtained when operating MILS in a swept-frequency regime instead of a constant frequency. Each frequency would provide $\Delta\varphi$ and $P/P_{vac}$ values, and such an increase in the number of output quantities can improve the density reconstruction accuracy, similarly to the installation of additional receivers. The practical realization can be limited by the width of the frequency band supported by a waveguide, and the profit of such operation regime, relative to the drawback of its relative complexity for the analysis, needs to be assessed and it is a work in progress.

Non-monotonic behaviour of the phase and power in the parameter space consitutes a difficulty in the density profile reconstruction, as can be seen, for example, for profile #7 in Fig. 14b. Such non-monotonic regions are present in the current database, but they are fortunately either small or affect only densities which are of less interest for practical applications on ASDEX Upgrade. This complication could be overcome either with an additional receiver or with other alterations to the diagnostic.

The chosen parameters of the MILS diagnostic (frequency, sizes, location, etc.) result in excellent coverage of a large range of densities. By adjusting these parameters, the density range (and the covered radial location) can be changed. For instance, shortening the distance between the horns, or their retraction away from plasma would lead to smaller range of radial coordinate of the detected density, as well as would change phase and power distribution in the parameter space, thereby giving better resolution to the high-density part and excluding some low densities from the detection range. By such adjustments, the diagnostic can be fine-tuned to provide even higher accuracy of density reconstruction in a target range of $n_e$.

### 9.3 Experimental factors

In practice, an additional error might come from a misalignment of the inteferometer axis relative to the background magnetic field. It is important to not allow any significant fraction of the power propagating as the X-mode, since it would disturb the O-mode measurements.

Another practical aspect is the density poloidal homogeinety. Large density perturbations as filaments or ELMs should be analysed separately, but even a misalignment of the plasma flux surfaces with the limiter causes a deviation fom the assumptions made in this paper. Small differences of the plasma shape should not play a significant role, but exotic shapes with strong misalignment cannot be processed in the framework of the constructed database and require a dedicated modelling.

Some level of turbulence is always present in realistic plasmas (even when not considering largely inhomogeneous density profiles with ELMs) and therefore it can never be described by a perfectly smooth profile in all three dimensions. In this context, the measurements of MILS can be considered as averaged over the low levels of turbulence. Unlike the microwave reflectometers, for which it is known that the turbulence can represent a problem for the possibility of measurements (mostly due to strong Bragg back-scattering [25]), in our region of interest the density is substantially lower than the cut-off density, therefore, we can consider all wave scattering events as negligible. Regarding the refraction of the MILS beam, used as one of the main principles of our diagnostic technique, the power level of the signal cannot be impacted significantly by low level turbulence, since the refraction index $N$ changes significantly only at values $\sim 0.5*n_c \approx 1.4*10^{19}$ m$^{-3}$ (Fig. 15), while all our measurements are far below this value.

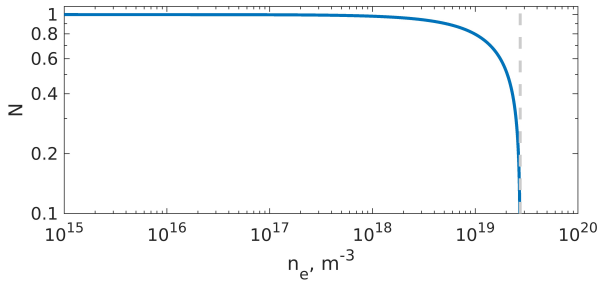

Fig. 15. Dependence of the refractive index on density. Gray dashed line shows the cut-off density $n_c = 2.74*10^{19}$ m$^{-3}$.

There is a large range of experimental conditions, in which the turbulence can be considered low. In [26], fluctuations of density in SOL on the level of 20-80 % are reported from experimental measurements on Alcator C-mod. Much stronger turbulence, with the level that changes the order of magnitude of the density in the measurement region (and leads to a corresponding drastic change in $N$), would be required in order to affect significantly the power of MILS signal. In such conditions the procedure of data analysis, presented in this paper, should not be applied.

In all other conditions, we stay in the regime when the deviations of MILS power and phase signals scale linearly with the density fluctuations. This justifies the possibility of considering the averaged signal. In principle, the linearity or non-linearity of the regime can be revealed for a given dataset from the time evolution of the signal. In the non-linear case, the dataset can be considered as not suitable to be processed with the pesented data analysis method. In the linear case, the signal deviation from the average over time has to be accounted for when evaluating the errorbars. The total error on the reconstructed density profile would consist of the sum of the numerical method errors and the experimental errors, the latter one expected to be larger than the former one in majority of the cases.

While the averaged signals are of the major interest, the diagnostic can be also used to study individual and statistical properties of SOL plasma density – turbulence, ELMs and filaments. Large density perturbations should dominate the MILS signals, when present in the measurement region; they can be well resolved temporally and absolute values of the density perturbation could be estimated. A separate analysis is ongoing for this topic, with some reslults available in [27].

One of the practical advantages of MILS compared to a reflectometer is the configuration of its antennnas. They are not directed towards plasma, but tangentially to it, which reduces drastically the stray radiation perturbations usually present in microwave diagnostics.

### 9.4 Other applications

The current study has demonstrated the application of the new density measurement technique for the region of far SOL, but the same principle can be applied to measurements further towards higher densities. Larger wave frequency and a modified location of horn antennas would enable measurements of other areas of edge plasma.

Due to the very high temporal resolution of MILS and the robustness of components, needed for its practical implementation, the diagnostic can be an attractive option for a plasma position control tool in DEMO. The main currently considered option, the magnetic detectors, do not yet demonstrate results, which could assure the controllability of the DEMO plasma [28]. Microwave disagnostics, such as reflectometry and MILS, need to be investigated in details to provide a reliable solution for a reactor.

### 10. Conclusions

Characterization of far-SOL density in tokamaks is usually not targeted specifically and is rather done together with measurements in near-SOL. The diagnostics used for it are therefore not well adjusted for the range of densities of $10^{15}$–$10^{18}$ m$^{-3}$. As a result, systematic information on the typical density profiles in that region is not available, though it can be very valuable for studies in the areas of plasma-wall interactions, ICRF and Lower-Hybrid heating, etc. Reflectometric diagnostics can also profit from the density data at the very edge, since the determination of exact radial location of their measurements is hindered by the difficulty of first fringe detection [13].

A new diagnostic technique – MILS – is presented here as a concept and analysed with parameters chosen for the ASDEX Upgrade edge conditions. The diagnostic is non-intrusive and the physical principle is simple. The reconstruction of radial density profile is done from the values of measured phase shift and power alteration of the wave. 3D forward modelling is performed with precise interferometer geometry and full-wave physics description, and it is optimized to carry out large parametric scans of input density profiles in reasonable computational time, hence giving a possibility to construct a sufficiently large database. In combination with the developed genetic algorithm genMILS of density profile reconstruction from the built database, it result in a powerful tool for MILS data interpretation. Common methods, used for microwave diagnostics modelling, such as ray-tracing or 2D full-wave codes, would not be able to achieve comparable accuracy. Therefore, the developed numerical tools play a crucial role in the demonstration of the capability of the presented novel diagnostic technique to provide far-SOL density estimation in a broad range of densities (~$10^{15}$–$10^{19}$ m$^{-3}$ with the considered MILS parameters) with high accuracy.

The numerical error of the presented method of density reconstruction, within the applied numerical algorithm and for the chosen MILS parameters (without being generalized to the diagnostic technique in general), is estimated as 5–15 % for $n_e \geq 10^{17}$ m$^{-3}$, it can be the same or higher for $n_e = 10^{16}$–$10^{17}$ m$^{-3}$, and for $n_e \leq 10^{16}$ m$^{-3}$ it is expected that the order of the density value can be obtained. In application to experimental data, this error is only a part of the total error, which also includes the experimental uncertainties (cables noise, thermal drifts, data acquisition errors, plasma turbulence level, etc).

On of the major advantages of the proposed diagnostic technique is that the temporal resolution can be chosen as high as an order of magnitude below the probing frequency (~ GHz) allowing for various studies of density changes at different time scales. MILS concept is applicable not only to tokamaks, but to stellarators and other devices as well, as long as a good alignment perpendicular to the background magnetic field for O-mode propagation can be provided. There are several possible modifications of the diagnostic, suggested in the paper, such as usage of more receivers or alterations of MILS parameters, which could increase the accuracy of the reconstruction of the density profile or allow changes in the coverage of the density range and/or radial location. Near SOL densities of a tokamak could be diagnosed with modified MILS.

The interpretative modelling is planned to be further advanced. The accuracy of inter-ELM like plasma profile reconstruction, presented in this work, can be increased by enlarging the number of parameters, used to describe the density profile shape. On the other side, non-monotonic 2D or 3D profiles, for example with ELMs or with turbulence, can be modelled using the developed synthetic diagnostic and this transient events can be studied. Experimental density profiles can be used as input in the developed synthetic diagnostic and numerical results can be compared to corresponding experimental measurements.

The presented method of data interpretation is ready to be applied to the processing of experimental data of MILS diagnostic, installed on the ASDEX Upgrade tokamak. The question of validity of the diagnostic technique and of the developed approach of data interpretation can be influenced by exact experimental conditions. It is out of the scope of the current study to predict all possible experimental limitations; they have to be accounted for during practical applications. Large variations by orders of magnitude of the far SOL density in L- and H-modes in modern tokamaks, with values on the edge of or below the detection limits of other diagnostics, provide conditions where MILS can give crucial contribution to the exsisting knowledge.

## Acknowledgements

The authors wish to thank V. Bobkov, B. Tal and G. Urbanczyk for helpful discussions on the topic of edge plasma density measurements.

This work has been carried out within the framework of the French Federation for Magnetic Fusion Studies (FR-FCM) and EUROfusion Consortium and has received funding from the Euratom research and training programme 2014–2018 and 2019–2020 under Grant agreement No. 633053. The views and opinions expressed herein do not necessarily reflect those of the European Commission.

## Availability of data

The data that support the findings of this study are available from the corresponding author upon reasonable request.

## Author Declarations

The authors have no conflicts to disclose.

## References

[1] Fessey J A *et al* 1987 *J. Phys. E: Sci. Instrum.* **20** 169
[2] James R A *et al* 1995 *Rev. Sci. Instrum.* **66** 422
[3] Varavin M *et al* 2014 *Telecommun.Radio Eng.* **73** 571
[4] Varavin M *et al* 2019 *Fus. Eng. Des.* **146 B** 1858
[5] Nam Y U *et al* 2008 *Rev. Sci. Instrum.* **79**, 10E705
[6] Su P *et al* 2021 *Rev. Sci. Instrum.* **92**, 043538
[7] Van Milligen B Ph *et al* 2011 *Rev. Sci. Instrum.* **82**, 073503
[8] Filippov V *et al* 2020 *Rev. Sci. Instrum.* **91**, 093503
[9] Pavlichenko R O *et al* 2018 *Probl. At. Sci. Technol.* **6**, 118
[10] Usoltceva M *et al* 2018 *Rev.Sci. Instrum.* **89** 10J124
[11] Boxer A C *et al* 2009 *Rev.Sci. Instrum.* **80**, 043502
[12] Dittmann K *et al* 2012 *Plasma Sources Sci. Technol.* **21** 024001
[13] Aguiam D E *et al* 2016 *Rev.Sci. Instrum.* **87**, 11E722
[14] Usoltceva M *et al* "Experimental results of MILS measurements of far-SOL density", *in preparation*
[15] Jacquot J *et al* 2013 *Plasma Phys. Control. Fusion* **55** 115004
[16] Tierens W *et al* 2019 *Nucl. Fusion* **59** 046001
[17] Bobkov V *et al* 2021 *Nucl. Fusion* **61** 046039
[18] Usoltceva M *et al* 2021 *Fus. Eng. Des.* **165** 112269
[19] Tierens W *et al* 2020 *AIP Conf. Proc.* **2254** 070005
[20] Suárez López G *et al* 2020 *Plasma Phys. Control. Fusion .* **62** 125021
[21] Stix T H 1992 Waves in plasma *AIP*, second edition
[22] Stangeby P C 2000 The Plasma Boundary of Magnetic Fusion Devices *Bristol: Institute of Physics Publishing*
[23] Yang X-S 2021 Nature-Inspired Optimization Algorithms *Academic Press*, second edition, Chapter 6
[24] Claeskens G and Hjort N L 2008 Model Selection and Model Averaging *Cambridge University Press*, Chapter 7
[25] Zadvitskiy G V *et al* 2018 *Plasma Phys. Control. Fusion* **60** 025025
[26] Terry J L *et al* 2001 *J. Nucl. Mater.* **290-293** 757-762

[27] Usoltceva M *et al* "Sensitivity of Microwave Interferometer in the Limiter Shadow to filaments in ASDEX Upgrade", http://arxiv.org/abs/2110.01314

[28] Ariola M *et al* 2019 *Fus. Eng. Des.* **146A** 728-731

## Appendix A. Model optimization

*A.1 Initial model*

Finite element modeling is based on dividing the whole simulation domain into smaller elements, by constructing a mesh in 3D (or 1D, 2D) space, fine enough to acurately follow details of geometry shape. For each mesh cell, a solution of a set of equations is found and together with solutions for all other elements it constitutes an approximation of the solution for the whole modelled problem. The element order, or discretization, corresponds to the order of the equations being solved for each mesh cell. In COMSOL, the default recommended order is quadratic for partial differential equations, which have a dominant second derivative term.

Increasing discretization, while keeping the same mesh, leads to a similar effect as mesh refinement at constant element order, to the improved accuracy of the model. Mesh refinement can be more beneficial when the model has parts of very different sizes and requires complicated meshes of several pieces with different settings. For more or less homogenous models, the element order increase might be more favorable.

In this modelling, a combination of both the discretization increase and the mesh settings variation has been applied, in order to find the best possible model with highest precision and minimum resources consumption.

The starting point was chosen according to COMSOL recommendations: quadratic element order and mesh with 5 elements per wavelength in the direction of wave propagation (mesh element size of $l_{mesh\ y}$ = 6.4 mm / 5 = 1.28 mm). Due to much lower field variation in the direction perpendicular to MILS axis, a scaling was applied to the mesh along both $x$ and $z$ axes with a coefficient $s_{mesh}$ = 0.33, which means that the mesh element size was 3 times bigger in those directions: $l_{mesh\ x} = l_{mesh\ z}$ = 3.83 mm. This way the number of mesh elements was reduced approximately by $3^2$ times. The resulting mesh contained $N_{mesh}$ = 10*10$^6$ elements.

The details of this initial version of model and of all other versions are summarized in Table 3. Note that of the total number of elements, most are located in plasma and PML domains, while the vacuum domain, corresponding to the horn anetnnas and waveguides, contains 1–2 order of magnitude less elements. Therefore the details of its mesh have much less influence on the model optimization, and the only important thing to mention is that this domain had ~2–3 times finer resolution than the other domains, due to smaller geometry elements and the importance of fine resolution at the location where the output values are measured. The shape of all mesh elements in all domains was tetrahedral.

*A.2 Mesh element order (discretization) and size*

Unfortunately, the performance of the first version of the model was not satisfactory. Such a large model could not be solved on the available cluster (3 nodes * 1 TB, 48 cores each), unless going into the out-of-core calculation mode. The duration of such caclulation is extremely long and moreover unpredictable, due to the communication of the program with the hard drive memory instead of RAM.

One possible way to reduce the model size was by simultaneously increasing the discretization to cubic and making the mesh coarser. The mesh element size can be enlarged as long as the same precision of the solution can be kept. The size was increased equally in all 3 directions by multiplying to a coefficient $c_{mesh}$. Fig. 16 shows the mesh convergence study for 3 different cases of plasma density profile. Coefficients $c_{mesh}$ = 2, 1.9, 1.7, 1.5, 1.35 resulted in the number of mesh elements, correspondingly, $N_{mesh}$ = 1.2*10$^6$, 1.4*10$^6$, 2*10$^6$, 2.8*10$^6$, 4.1*10$^6$. The $y$ axis shows the deviation of the $\Delta\varphi$ value in the cubic discretization model from the value obtained in the model with quadratic discretization, described above and having $N_{mesh}$ = 10*10$^6$.

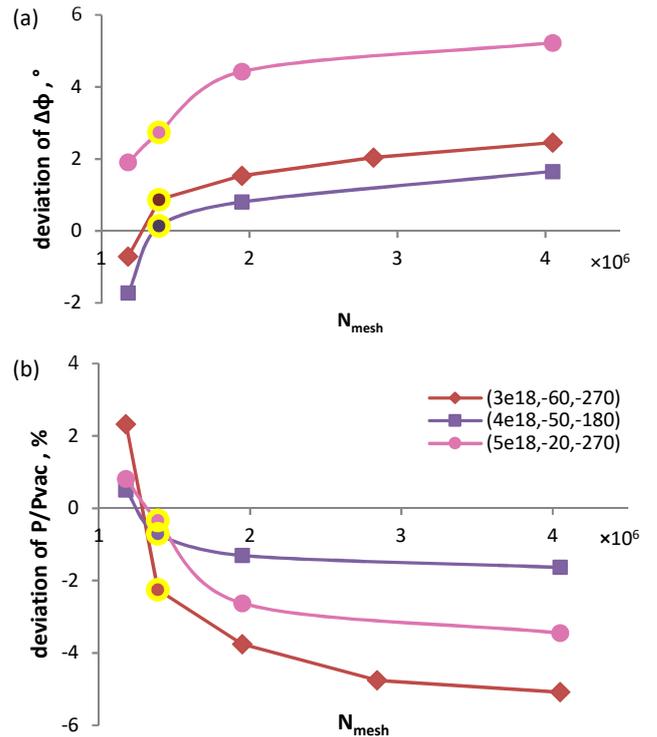

Fig. 16. Mesh convergence study for 3 different plasma profiles specified as $(n_{lim}, a_{in}, a_{out})$ set of parameters. Deviation of (a) phase and (b) power is calculated relative to the quadratic discretization model. Yellow markers show the model, taken for the next optimization step.

It can be seen for all 3 cases, that the cubic discretization results converge to values higher than the values obtained from the quadratic discretization model. The employed quadratic model shows visibly worse resolution than the most precise cubic model used. Therefore the general conclusion is that lower numerical resolution introduces a systematic error that leads to $\Delta\varphi$ underestimation. From the $c_{mesh}$ scan, a model should be chosen, which gives results close enough to the converged value and does not require too much resources. $c_{mesh} = 1.9$ ($N_{mesh} = 1.4*10^6$) is taken for further analysis, which is at least not worse than the quadratic model, and it shows relatively fast performance (at $c_{mesh} = 1.7$ it is already 2 times slower and requires 2 times more RAM, which is close to 3 TB limit of the available cluster). The chosen model is included in Table 3 as #2.

*A.3 Size of plasma domain*

In both the 1st and the 2nd model versions, the size of the plasma was kept the same. In order to save the computational resources, the minumum possible size is desired. While the size in $y$ direction is defined by the horn antennas positon and is therefore fixed, the sizes in the other 2 directions could be altered. There are two constraints for reducing the plasma size: possible reflections from the sides and inclusion of the whole region where the wave can propagate and be collected by the receiver.

The PML is best suited for absorbtion of the perpendicularly incident waves and its reflection coefficient reaches 1 fot tangential incidence. In our model, it can be seen that the wave reaches the PMLs in $x$ and $z$ directions at rather small angles. Reflections occur and unphysical negative power dissipation takes place in these PMLs. It has been investigated, to what extent this can affect the values, measured in the model. The plasma sizes are defined as the distance from the MILS axes to the $z$ direction $l_z$ (toroidal), to the positive $x$ direction $l_{x+}$ (towards the tokamak wall) and to the negative $x$ direction $l_{x-}$ (towards the plasma core).

In Fig. 17, the results of the analysis for one density profile (1e18,-50,-200) are presented, for 11 different cases of plasma sizes, given in the table (in cm). The biggest model #0 is used as a reference and it can be seen in Fig. 18 on the plot of Poynting vector main component that very small amount of power, going from the emitter antenna (lower horn), is directed towards the side PMLs in such a large model. Quantitatively, it can be characterized by the ratio between negative (unphysical) and positive (correct) power $P^-/P^+$ dissipated in PML, which for models #0 is quite low, 4.7 %. The maximum error in the measured quantites can be the square root of this value, and only if the whole power, reflected from the side PMLs, reaches the receiver, which is far from possible. The values plotted in Fig. 17 are calculated relative to this reference model.

It turns out that the plasma size can be reduced considerably, without significantly loosing the precision of measured quantities. The smallest model with not too large deviations of power and phase is #8, indicated in yellow in Fig. 17. This model is taken for further simulations. It has been checked at several examples of plasma density profile that with this plasma size the error in phase and power remains low (see Table 2). Note that the last 3 models in the table should be calculated with the model #6, with larger size, to keep the error low, as will be discussed below. Therefore these three models have larger errors.

| $n_e$ profile: $(a_{in}, a_{out}, n_{lim})$ | $\Delta\varphi$ deviation, ° | $P/P_{vac}$ deviation, % |
|---|---|---|
| (10e18,-10,-230) | 1,13 | -2,6 |
| (3e18,-20,-250) | 0,54 | -1,2 |
| (1e18,-50,-200) | 0,95 | -1,4 |
| (0.9e18,-80,-240) | 0,62 | 0,7 |
| (0.6e18,-60,-320) | 0,27 | 3,0 |
| (0.3e18,-80,-280) | 0,54 | -1,5 |
| (0.1e18,-100,-240) | -2,47 | -0,6 |

Table 2. Deviation of phase and power in models with plasma size #8, relative to a reference model #0, for several density profiles. Some cases are shown in gray, because the used plasma size is too small for them, so they should not be considered for the conclusions.

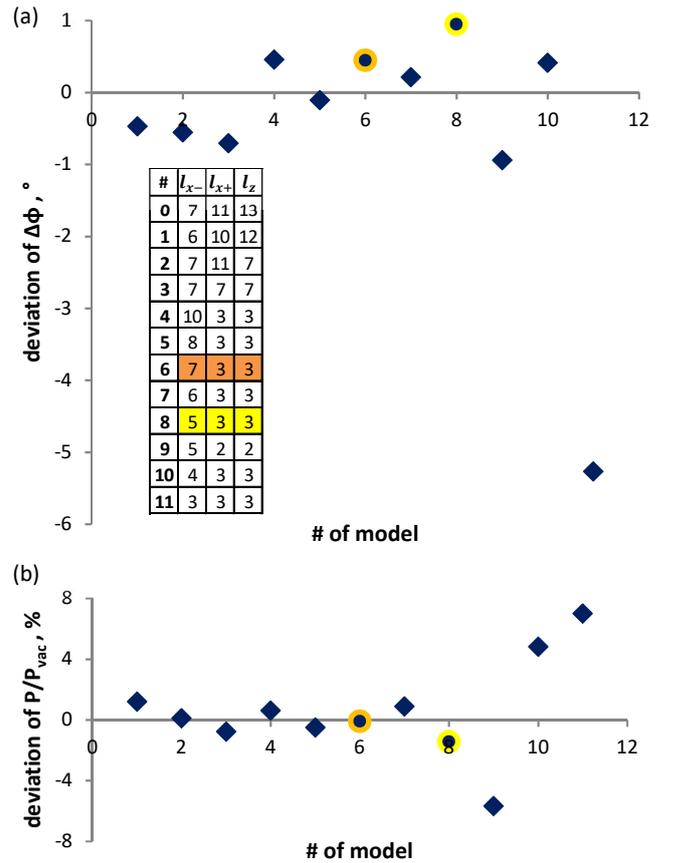

Fig. 17. Deviation of (a) phase and (b) power in models with reduced plasma size, relative to a reference model #0. Plasma sizes (see Fig. 18) are given in the table, in cm. Orange and yellow markers indicate models, chosen for further studies.

In plasma with lower density, the wave propagates further in the negative $x$ direction and its refraction at larger distance becomes important. In order to account for this constraint on the plasma size, a bigger model is needed, and it was decided to utilize the model #6, indicated in orange in Fig. 17. Both models are included as model version 3 in Table 3.

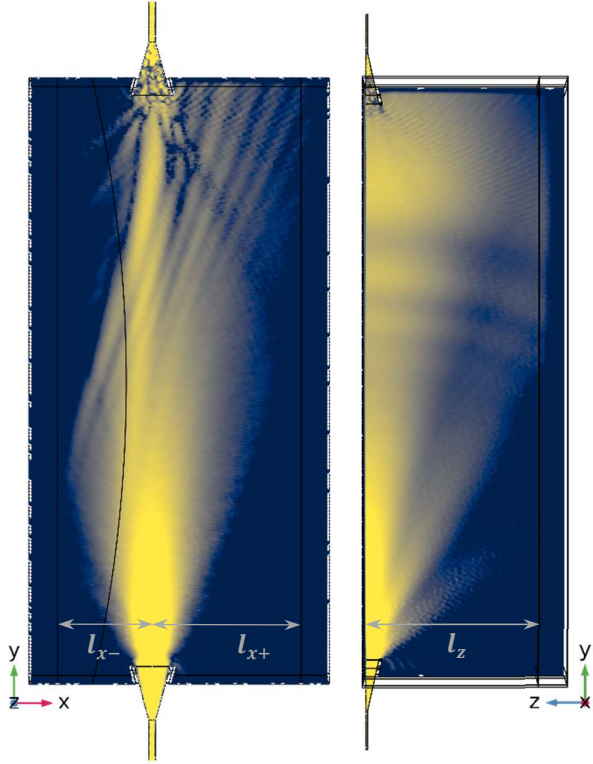

Fig. 18. Poynting vector in log10 scale, $y$-component. Model #0 from Fig. 17, with plasma profile (1e18,-50,-200). Left: projection on $x$-$y$ plane at $z = 0$; right: projeciton on $z$-$y$ plane at $x = 0$.

The separation of parameter space in 2 parts, models with $l_{x-} = 5$ cm and models with $l_{x-} = 7$ cm, is done based on the value of the largest density in the whole domain, at $x = -5$ cm in the middle along $y$. This value is representative in assessing the refraction of the interferometer wave by the plasma. In Fig. 19a the distribution of this value in the parameter space $(n_{lim}, a_{in})$ is shown (no dependence on $a_{out}$). In Fig. 19b,c the results of comparison of several density examples are given, indicating the error in phase and power, when a model with $l_{x-} = 5$ cm is used instead of $l_{x-} = 7$ cm for lower densities. Based on this analysis, a boundary was set at $n_e(x = -5 \text{ cm}) = 5.62*10^{18}$ m$^{-3}$, and models with density profiles, for which this value is below or equal to this boundary, were simulated with enlarged $l_{x-} = 7$ cm. The boundary is shown as dashed line in Fig. 19.

Such an effort to have 2 model versions is motivated by the need to speed up the calculations. The smaller model with $l_{x-} = 5$ cm is ~25 % faster that the bigger one, which results in noticeable computational time difference for large parameter scans.

*A.4 Improved accuracy with modified mesh scale*

The drastic reduction of the computational time and RAM, achieved at the previous optimization step, allowed more freedom in balancing the model accuracy and the computational cost. At this step, we have scanned the mesh scale $s_{mesh}$ value, to make sure that it was chosen adequately. Fig. 20 demonstrates results of a scan with $s_{mesh}$ = 0.33, 0.42, 0.5, 0.75, 1 in a model with all other settings the same. Three different density profiles were used. The values are calculated relative to the model with highest resolution, with $s_{mesh} = 1$. Based on this result, it was decided to choose $s_{mesh} = 0.42$ instead of previously used 0.33 and therefore achieve better precision. Higher $s_{mesh}$ values were not chosen because of too large increase in the computational cost ($s_{mesh} = 0.5$ is nearly 2 times slower than $s_{mesh} = 0.42$ and needs nearly 2 times more RAM).

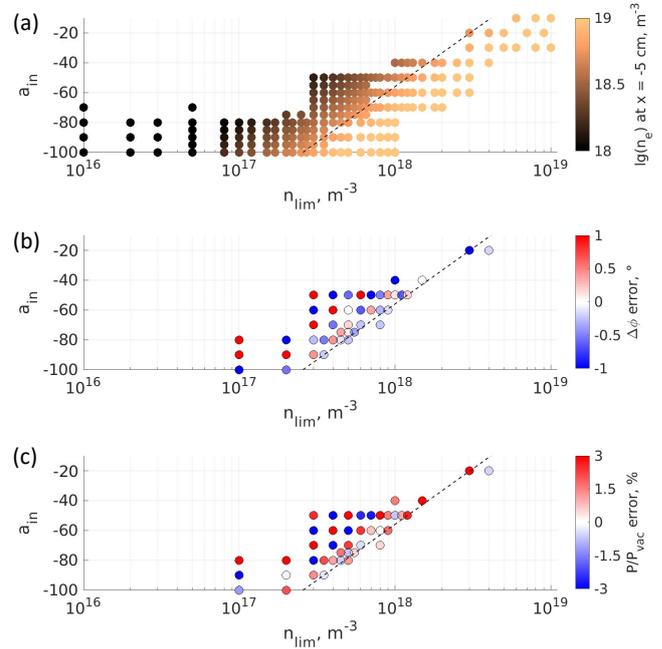

Fig. 19. (a) Denisty at x = -5 cm in the middle of MILS axis height, as a function of $(n_{lim}, a_{in})$; (b) phase and (c) power error in models with $l_{x-} = 5$ cm compared to models with $l_{x-} = 7$ cm. Dashed lines show the boundary used to separate the parameters space for the usage of smaller or bigger model.

The settings of this model version are given in Table 3 as number 4. This model was used to produce all results with plasma density profiles, presented in the main text of this paper.

*A.5 Error assessment*

The errors in the final model, caused by numerical imprecision, can be summarized as:
- underestimation of $\Delta\varphi$ by 1.5−2.5° because of mesh size increase relative to the optimal (given by $c_{mesh}$);

- overestimation of $\Delta\varphi$ by ~ 1° because of plasma size reduction relative to the optimal;
- underestimation of $\Delta\varphi$ by ~ 2° because of mesh size scaling perpendicularly to MILS axis, relative to the optimal (given by $s_{mesh}$);
- overestimation of $P/P_{vac}$ by 1−3 % because of mesh size increase relative to the optimal (given by $c_{mesh}$);
- deviation of $P/P_{vac}$ by ~ -2.5−1 % because of plasma size reduction relative to the optimal;
- overestimation of $P/P_{vac}$ by ~ 2 % because of mesh size scaling perpendicularly to MILS axis, relative to the optimal (given by $s_{mesh}$).

The mesh size change by the variation of $c_{mesh}$ value corresponds to mesh refinement in all 3 directions, while the change of $s_{mesh}$ results in the mesh alteration along 2 out of 3 directions. The influence of these 2 ways of mesh adjustment on the model output is observed to be similar and therefore the errors should not be added up, but only one of them (the latter) should be taken. The plasma size reduction introduces an error, which has a different origin (reflection from the walls), hence this error should be summed with the error from the mesh, to get the full error. It gives systematic errors $\delta_{\varphi\,sys}$ = -1° and $\delta_{P\,sys}$ = +1 % and random errors $\delta_{\varphi\,rnd}$ < 0.5° and $\delta_{P\,rnd}$ = 1.5 %.

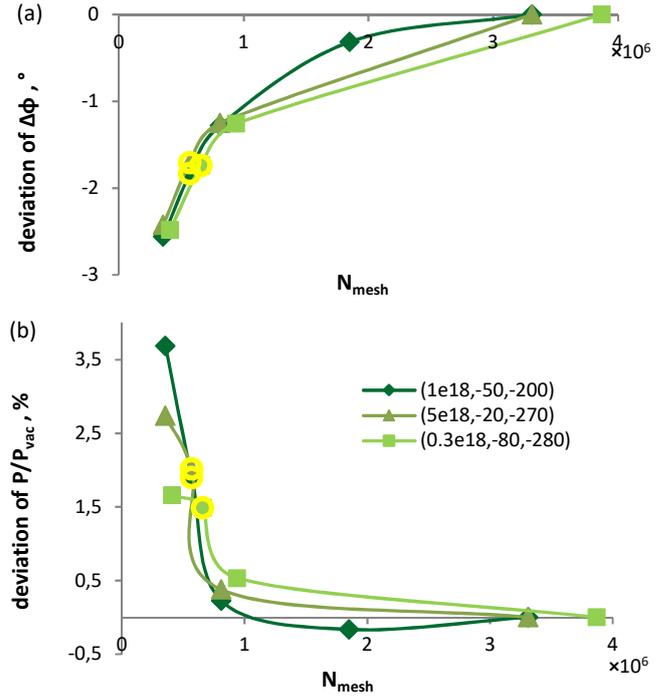

Fig. 20. Deviation of (a) phase and (b) power depending on mesh scale factor $s_{mesh}$, as a function of the total mesh elements number. Yellow marker indicates the chosen model. Density profile (1e18,-50,-200).

| Model version | Discretization | $l_{mesh\,y}$, mm | $s_{mesh}$ | Plasma sizes $d_z$, $d_{x+}$, $d_{x-}$, cm | # of mesh elements | DoF; RAM used | Solution time (# of nodes) |
|---|---|---|---|---|---|---|---|
| 1 | Quadratic | $\lambda/5$ = 1.28 | 0.33 | 10, 10, 6 | $10*10^6$ | $64*10^6$; >3 TB | 8-20 h (3), out-of-core |
| 2 | Cubic | $\lambda/5*1.9$ = 2.42 | 0.33 | 10, 10, 6 | $1.4*10^6$ | $26*10^6$; 1.5 TB | 1 h (3) |
| 3 | Cubic | $\lambda/5*1.9$ = 2.42 | 0.33 | 5, 3, 3 or 7, 3, 3 | $0.36*10^6$ $0.41*10^6$ | $6.7*10^6$, 0.15 TB $7.8*10^6$, 0.18 TB | 10 min (1) 13 min (1) |
| 4 | Cubic | $\lambda/5*1.9$ = 2.42 | 0.42 | 5, 3, 3 or 7, 3, 3 | $0.57*10^6$ $0.66*10^6$ | $10.8*10^6$, 0.29 TB $12.4*10^6$, 0.4 TB | 23 min (1) 29 min (1) |

Table 3. Summary of model versions, used at different optimization steps. The last one is used to produce results presented in the main article text.